\documentclass[%
aip,
cha, 
reprint,%
floatfix,%
superscriptaddress]{revtex4-2}

\usepackage{amsmath,amsthm,amssymb,amsfonts}

\usepackage{setspace}
\usepackage{mathrsfs}
\usepackage{cases}

\usepackage{graphicx}
\usepackage{dcolumn}
\usepackage{bm}

\usepackage{etoolbox}

\usepackage[usenames]{xcolor}

\makeatletter
\def\@email#1#2{%
 \endgroup
 \patchcmd{\titleblock@produce}
  {\frontmatter@RRAPformat}
  {\frontmatter@RRAPformat{\produce@RRAP{*#1\href{mailto:#2}{#2}}}\frontmatter@RRAPformat}
  {}{}
}%
\makeatother

\numberwithin{equation}{section}

\newcommand{\BbbR}{\mathbb{R}}

\DeclareMathOperator{\si}{si}

\newcommand{\scrI}{\mathscr{I}}

\begin{document}

\title{Quantum kicks near a Cauchy horizon}

\author{Benito A. Ju\'arez-Aubry}
\affiliation{Instituto  de  Ciencias  Nucleares,  
Universidad  Nacional Aut\'onoma de M\'exico, 
Mexico City 045010, Mexico}\thanks{Present address.}
\affiliation{School of Mathematical Sciences, University of Nottingham, Nottingham NG7 2RD, UK}
\author{Jorma Louko}%
\affiliation{School of Mathematical Sciences, University of Nottingham, Nottingham NG7 2RD, UK}%
\email{benito.juarez@correo.nucleares.unam.mx, jorma.louko@nottingham.ac.uk}

\date{September 2021; revised December 2021}

\begin{abstract}
We analyse a quantum observer who falls geodesically towards 
the Cauchy horizon of a $(1+1)$-dimensional eternal black hole 
spacetime with the global structure of the non-extremal Reissner-Nordstr\"om solution. 
The observer interacts with a massless scalar field, using an Unruh-DeWitt detector coupled linearly to the proper time derivative of the field, and by measuring the local energy density of the field. Taking the field to be initially prepared in the Hartle-Hawking-Israel (HHI) state or the Unruh state, we find that both the detector's transition rate and the local energy density generically diverge on approaching the Cauchy horizon, respectively proportionally to the inverse and the inverse square of the proper time to the horizon, and in the Unruh state the divergences on approaching one of the branches of the Cauchy horizon are independent of the surface gravities. When the outer and inner horizons have equal surface gravities, the divergences disappear altogether in the HHI state and for one of the Cauchy horizon branches in the Unruh state. We conjecture, on grounds of comparison with the Rindler state in $1+1$ and $3+1$ Minkowski spacetimes, that similar properties hold in $3+1$ dimensions for a detector coupled linearly to the quantum field, but with a logarithmic rather than inverse power-law divergence. 
\end{abstract}

\maketitle

\section{Introduction}
\label{sec:intro}

It is a great pleasure to dedicate this paper to Roger Penrose, 
who realised the instability of the Cauchy horizons 
that occur inside charged and rotating black hole solutions~\cite{Simpson:1973ua}. 
The nature of the instability is a topic of ongoing research, 
classically and in the presence of quantised fields. 
This paper addresses transitions in a time-and-space localised quantum system, 
coupled to an ambient quantum field, 
when the system falls geodesically towards a Cauchy horizon. 

Causality is a concept at the core of physics. Classically, causality is formulated as the well-posedness of the initial value problem: a solution to the dynamical equations is fully determined by the initial conditions specified on a spacelike hypersurface. In the quantum theory, causality may be formulated in terms of an algebra of observables: the commutator of two observables whose respective supports have spacelike separation must vanish. In both cases, the physical meaning is that any two causally disconnected observables have no influence on each other.

In geometric terms, causality is protected by global hyperbolicity \cite{Hawking:1973uf,Wald:1984rg}. Every globally hyperbolic spacetime is stably causal, since global hyperbolicity implies the existence of a global time function that provides the stable causality condition. In turn, this implies strong causality, which prevents any causal curve from coming arbitrarily close to intersecting itself. Moreover, the dynamical equations of classical fields admit a well-posed initial value problem on the whole manifold whenever suitable data is specified on a Cauchy hypersurface of a globally hyperbolic manifold. 
With quantum fields, global hyperbolicity allows one to establish a rigorous quantisation scheme for free fields\cite{Wald:1995yp, Brunetti:2001dx}, which provides the starting point for a perturbative expansion in interacting theories\cite{Hollands:2014eia}. 

However, many important solutions in General Relativity are not globally hyperbolic: they contain Cauchy horizons, which are boundaries of the maximal Cauchy development of an achronal hypersurface. This includes most members of the analytically extended Kerr-Newman family. 

There is a significant history of work addressing the stability of Cauchy horizons in General Relativity. In the classical theory, work by Simpson and Penrose led to the \textit{strong cosmic censorship conjecture\/} \cite{Simpson:1973ua}, which states that for generic initial data the spacetime 
is inextendible beyond the maximal Cauchy development. Support for this conjecture came from Chandrasekhar and Hartle's observation that the (electromagnetic or gravitational) classical radiation felt by an observer diverges as the Reissner-Nordstr\"om horizon is approached\cite{Chandrasekhar:1982}. 
Later work has however revealed that the sense of inextendibility in the conjecture is subtle. 
On the one hand, 
given polynomially decaying initial data for the Einstein-Maxwell-scalar system, settling down to a Reissner-Nordstr\"om black hole, 
the spacetime is $C^0$-extendible past the Cauchy horizon; on the other hand, not 
all the geometric invariants remain finite and, 
in particular, the Hawking mass diverges at the Cauchy horizon. 
This is known as the mass inflation scenario \cite{Poisson:1989zz,Poisson:1990eh,Dafermos:2002ka,Dafermos:2003wr,Costa:2014zha}. 

When the theory is extended to include quantised fields, new issues arise from the renormalised stress-energy tensor near the Cauchy horizon, and from the back-reaction of this stress-energy on the spacetime. There is evidence that the back-reaction of the quantum fields tends to make Cauchy horizons generically unstable even in situations where classical surface gravity considerations would suggest stability \cite{Casals:2016odj,Casals:2019jfo,Hollands:2019whz,Hollands:2020qpe,Klein:2021les}. 

In this paper we shall address another facet of 
the singular behaviour of quantised fields near a Cauchy horizon: 
the experiences of 
a time-and-space localised quantum system as it falls geodesically towards the Cauchy horizon. 
Interaction with the ambient quantum field causes transitions between the internal states of the localised quantum system. 
Does the probability of these transitions change rapidly, perhaps even divergently, as the system approaches the Cauchy horizon? If so, is there a correlation between the rapid changes in transition probabilities and any divergent behaviour that the field's stress-energy tensor may exhibit near the Cauchy horizon? 

We shall consider a class of $(1+1)$-dimensional eternal black hole spacetimes whose global structure mimics that of the non-extemal Reissner-Nordstr\"om solution, 
with an asymptotically flat region, 
an outer bifurcate Killing horizon and an inner bifurcate Killing horizon \cite{Hawking:1973uf}, 
but allowing the `radial' profile function in the metric to remain otherwise arbitrary, 
and in particular allowing the outer and inner horizons to have arbitrary nonvanishing surface gravities. 
As the ambient quantum field, we consider a massless scalar field, prepared initially in the 
Hartle-Hawking-Israel (HHI) state \cite{Hartle:1976tp,Israel:1976ur} or in the Unruh state \cite{Unruh:1976db}. 
As the local quantum system, we consider a spatially pointlike two-level system known as the Unruh-DeWitt detector \cite{Unruh:1976db,DeWitt:1979}, in a variant that couples linearly to the proper time derivative of the field. The reason to include the derivative is that this makes the detector's transition probabilities independent of the scalar field's infrared ambiguity. 

We work within first-order perturbation theory. We assume the detector to be switched on and off instantaneously, and we address not the transition probability itself but the transition rate, defined as the derivative of the transition probability with respect to the switch-off proper time. 
While this amounts to ignoring a technically divergent `additive constant' 
contribution to the transition probability from the instantaneous switching\cite{Schlicht:2003iy,Louko:2006zv,Satz:2006kb,Obadia:2007qf,Louko:2007mu,Juarez-Aubry:2014jba}, 
it allows us to isolate the singular effects due to the approach to the Cauchy horizon, 
which effects are the focus of this paper. 

We find that as the geodesic detector approaches the Cauchy horizon, 
the transition rate diverges whenever the outer and inner horizons have differing surface gravities, 
on all parts of the Cauchy horizon. In the exceptional case of equal surface gravities, 
the transition rate remains bounded in the HHI state on all parts of the Cauchy horizon, 
and in the Unruh state on the branch of the Cauchy horizon that is opposite to the exterior 
with respect to which the Unruh state is defined. 
When the divergence occurs, it is proportional to the inverse 
of the proper time separation from the Cauchy horizon, except that in the Unruh state, 
for a geodesic approaching the Cauchy horizon bifurcation point, 
the divergence is slightly weaker when the outer horizon has 
twice the surface gravity of the inner horizon. 

We also find that these results for the transition rate are in significant qualitative and quantitative agreement with the divergences in the energy density seen by an observer on the geodesics. The main difference is that the energy density generically diverges proportionally to the inverse square, rather than the inverse, 
of the proper time separation from the Cauchy horizon; however, in the energy density averaged over the trajectory, the divergence is again proportional to the inverse of the proper time separation from the Cauchy horizon. 
The divergence in the stress-energy tensor, 
including the special role of the equal surface gravity case therein, 
has been studied in the context of back-reaction, in both $1+1$ dimensions and in $3+1$ dimensions\cite{Hollands:2019whz,Hollands:2020qpe,Klein:2021les}. 

Finally, we perform a similar analysis for a geodesic detector approaching the Rindler horizon in $(1+1)$-dimensional Minkowski spacetime, with the field prepared in the Rindler vacuum, and we contrast the results with a similar analysis in $3+1$ dimensions \cite{Louko:2007mu}, for a detector coupled linearly to the value (as opposed to the derivative) of the scalar field. Based on this comparison, we conjecture that in $3+1$ spacetime dimensions, 
a detector coupled linearly to the value of the scalar field, and approaching a Cauchy horizon, 
generically has a transition rate that diverges in proper time but only logarithmically.

We begin in Section \ref{sec:blackhole} by presenting our $(1+1)$-dimensional eternal black hole spacetime, 
discussing its similarities with the $(3+1)$-dimensional non-extremal Reissner-Nordstr\"om solution, 
presenting adapted coordinate systems, 
and recording properties of timelike geodesics that approach the Cauchy horizon. 
Section \ref{seq:quantum-scalar} introduces the massless scalar field and records its Wightman functions in the HHI and Unruh states. 
Section \ref{sec:detector} starts with a concise conceptual review of Unruh-DeWitt detectors as space-and-time localised quantum systems by which the quantum field is probed, specialises then to a detector whose coupling to the field includes a time derivative, and focuses finally on the detector's instantaneous transition rate, 
treated in first-order perturbation theory. 

Our main results, for the detector's transition rate on approaching the Cauchy horizon, are presented in Section~\ref{sec:detector-near}, deferring technical aspects to three appendices. 
Section \ref{sec:stress-energy} presents the corresponding results for the energy density on a geodesic, and 
Section \ref{sec:rindler} presents the comparison with the Rindler horizon, in $1+1$ and $3+1$ dimensions. 
Section \ref{sec:conclusions} gives a summary and concluding remarks. 

We use units in which $c = \hbar = k_B = 1$. 
In asymptotic expansions, 
${\cal O}(x)$ denotes a quantity such that ${\cal O}(x)/x$ is bounded as $x\to0$, 
$o(x)$ denotes a quantity such that $o(x)/x \to 0$ as $x\to0$, 
${\cal O}(1)$ denotes a quantity that is bounded in the limit under consideration, 
and 
$o(1)$ denotes a quantity that goes to zero in the limit under consideration. 

A subset of our results was announced previously in a conference proceedings contribution\cite{Juarez-Aubry:2015dla}. The key result given therein as formula (3.10), for the transition rate of a detector approaching the `left' branch of the Cauchy horizon when the field is in the HHI state, is our formula~\eqref{eq:fdot-final-hhi-LR}. The formulas have the same content, given the differing surface gravity conventions: 
in the present paper, the surface gravities of both inner and outer horizons are by definition positive, following the conventions of\cite{Hollands:2019whz,Hollands:2020qpe,Klein:2021les}, whereas in\cite{Juarez-Aubry:2015dla} the inner horizon surface gravity was defined to be negative. For the stress-energy, the present paper focuses on the energy density at a given moment on the trajectory, allowing a sharper asymptotic localisation than the time-averaged energy density discussed in Section 4 of\cite{Juarez-Aubry:2015dla}.

\section{The generalised Reissner-Nordstr\"om black hole in $1+1$ dimensions\label{sec:blackhole}} 

In this section we introduce a class of $(1+1)$-dimensional eternal black hole 
spacetimes that generalise the constant angles sections of the nonextremal Reissner-Nordstr\"om spacetime. 
We also write down the equations of geodesics approaching the Cauchy horizon in a convenient form.

\subsection{Metric and global structure}

Let $F: \BbbR^+ \to \BbbR$ be a smooth function 
such that 
\begin{align}
F(r_-) = F(r_+) = 0, 
\end{align}
where $r_\pm$ are constants satisfying $0< r_- < r_+$, 
\begin{subequations}
\begin{align}
&F(r) > 0\ \  \text{for}\ \ r_+ < r < \infty, \\
&F(r) < 0 \ \  \text{for}\ \ r_- < r < r_+, 
\end{align}
\end{subequations}
and 
\begin{subequations}
\begin{align}
F'(r_+) &= 2\kappa_+ > 0 ,  
\\
F'(r_-) &= - 2\kappa_- < 0 , 
\label{eq:innersurfacegrav}
\end{align}
\end{subequations}
where $\kappa_\pm$ are positive constants. 
We also assume that $F(r) \to 1$ as $r\to\infty$. 
Further information about $F(r)$ for $r<r_-$ will not be needed, but 
we note that it follows from the above that 
$F(r)>0$ when $r\in(a,r_-)$ for some $a<r_-$. 

To summarise, $F(r)\to1$ as $r\to\infty$, and $F$ has simple zeroes at $r=r_\pm$. 

We consider the spacetime metric 
\begin{align}
ds^2 = -F(r) dt^2 + \frac{dr^2}{F(r)} , 
\label{eq:RNcoords-metric}
\end{align}
where, to begin with, $r>r_+$. We refer to $(t,r)$ as Schwarzschild-like coordinates. 
This metric is static, with the timelike Killing vector~$\xi \doteq \partial_t$, 
and it is asymptotically flat at $r\to\infty$. 

\begin{figure}
\begin{center}
\includegraphics[width=0.99\columnwidth]{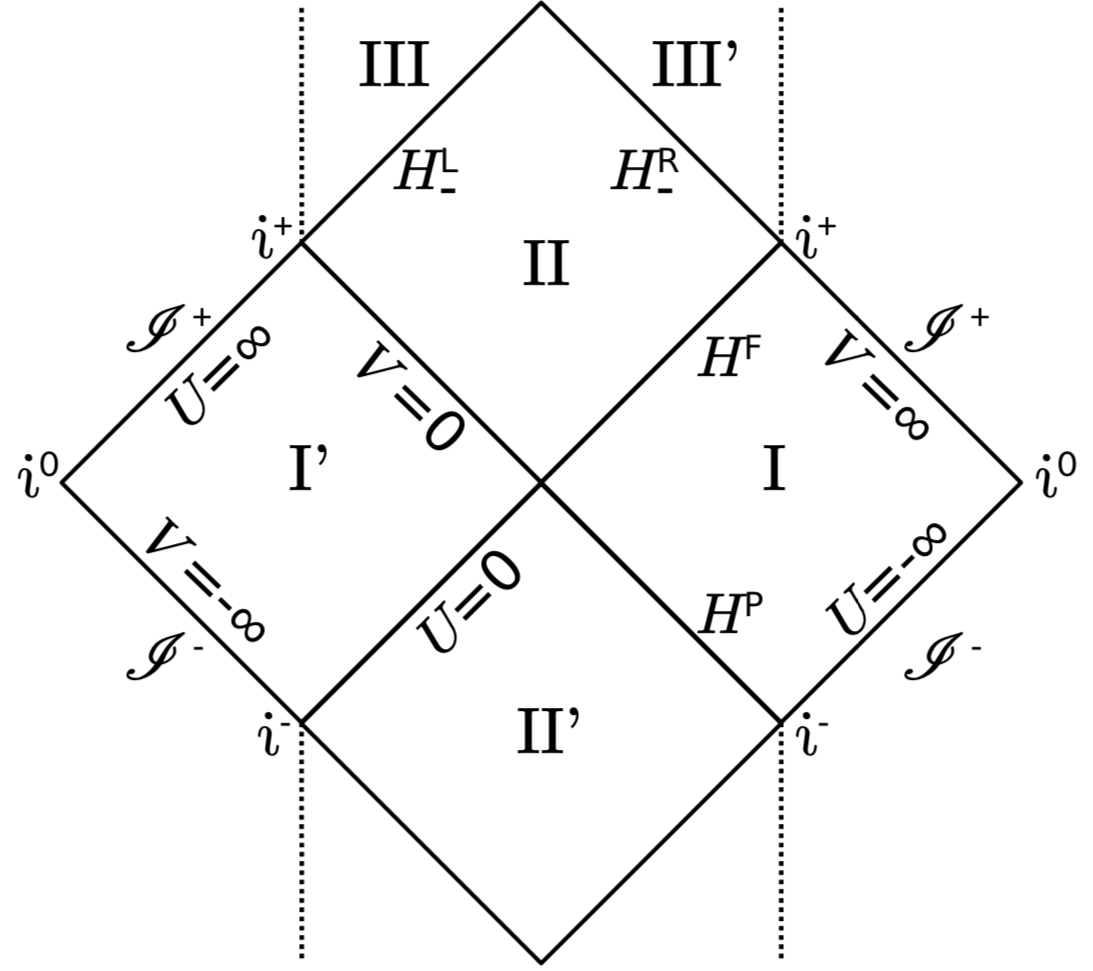}
\caption{Part of the conformal diagram of the extended spacetime. 
Region I is the `original' exterior~\eqref{eq:RNcoords-metric}, 
connected by the bifurcate Killing horizon at $r=r_+$ to the black hole interior~II, 
the white hole interior II' and the second exterior~I'. 
The Kruskal-like coordinates $(U,V)$ cover Regions I, II, II' and I', with the Killing horizon $r=r_+$ at $UV=0$. 
Regions II and II' are bounded in the future/past by the future/past Cauchy horizons at $r=r_-$.  
The dotted lines, bounding Regions III and~III, 
are singularities that occur behind the Cauchy horizons when $F(r)>0$ for $r<r_-$ and $F(r)\to\infty$ as $r\to0$; 
other structure behind the Cauchy horizons can occur under different behaviour of $F(r)$ for $r<r_-$. 
The diagram extends to the past and future.\label{RNConformal}}
\end{center}
\end{figure}

The metric has a smooth continuation across the coordinate singularity at $r=r_+$, 
and further a smooth continuation across the coordinate singularity at $r=r_-$. 
These continuations may 
be found by adapting the standard procedure for the Reissner-Nordstr\"om metric \cite{Hawking:1973uf}, 
for which $F(r) = (r-r_+)(r-r_-)/r^2$, 
and the continuations 
are real analytic when $F$ is real analytic. 
There is a bifurcate Killing horizon of $\xi$ at $r=r_+$, 
of surface gravity $\kappa_+$, 
and part of this Killing horizon forms the black hole event horizon 
with respect to the $\scrI^+$ of the original asymptotically flat region. 
On continuing to the past and to the future, there are further bifurcate Killing horizons of $\xi$ at $r=r_-$, 
of surface gravity $\kappa_-$, 
and they form past and future Cauchy horizons for the four regions joined by the original $r=r_+$ Killing horizon. 
The pattern continues to the past and to the future. What happens at $r<r_-$ depends on the behaviour of $F(r)$ there, 
and will not be needed here. 
The parts of the conformal diagram that are relevant for us are shown in Figure~\ref{RNConformal}, 
in the Reissner-Nordstr\"om-like case in which $F(r)>0$ for $r<r_-$ and $F(r)\to\infty$ as $r\to0$. 

\subsection{Coordinates}

We shall write down three coordinate systems that cover (at least) the original exterior 
region and the black hole interior region, and are adapted to 
the quantum states that we shall describe in Section~\ref{seq:quantum-scalar}.

\subsubsection{Kruskal-like coordinates}

We denote the `original' $r>r_+$ region of \eqref{eq:RNcoords-metric} by Region~I\null. 
In Region~I, define first the tortoise coordinate $r_* \in \BbbR$ by 
\begin{align}
dr_* = \frac{dr}{F(r)} ,
\label{eq:tortoise-def}
\end{align} 
making some arbitrary choice for the additive constant in~$r_*$, 
and then the Eddington-Finkelstein double null coordinates $(u,v) \in \BbbR^2$ by 
\begin{subequations}
\begin{align}
u &= t-r_* , 
\\
v &= t+r_* , 
\end{align}
\end{subequations}
and finally the Kruskal(-Szekeres)-like coordinates $(U,V) \in \BbbR^- \times \BbbR^+$ by 
\begin{subequations}
\begin{align}
U &= - e^{-\kappa_+ u}, 
\label{eq:EF-to-Kruskal-U}
\\
V &= e^{\kappa_+ v}. 
\label{eq:EF-to-Kruskal-V}
\end{align}
\end{subequations}
The metric takes the form 
\begin{equation}
ds^2 = \frac{F(r)}{\kappa_+^2 U V} \, dU \, dV, 
\label{eq:kruskalmetric}
\end{equation}
where $r$ is determined as a function of $U$ and $V$ from 
\begin{align}
-UV = e^{2\kappa_+ r_*}
\ . 
\label{eq:kruskalinversion}
\end{align}

\begin{table}
\begin{center}
\begin{tabular}{l | l l l l l l l l l }
& Region & $r$ domain &&& $U$ & $V$ &&& $\xi_a \xi^a$\\
\hline
I & Original exterior \ \ &$r_+ < r < \infty$ &&& $-$ & $+$ &&& $-$\\
II & Black hole &$r_-< r < r_+$ &&& $+$ & $+$ &&& $+$\\
II' & White hole &$r_-< r < r_+$ &&& $-$\ \  & $-$ &&& $+$\\
I' &Second exterior &$r_+ < r < \infty$ &&& $+$ & $-$  &&& $-$
\end{tabular}
\caption{The four subregions of the Kruskal-type chart $(U,V)$, in the labelling of Figure~\ref{RNConformal}. 
The last three columns indicate the signs of $U$ $V$ and $\xi_a \xi^a$ in each of the subregions.}
\label{tab:exterior}
\end{center}
\end{table}

It follows from the assumptions about $F$ that the metric 
given by \eqref{eq:kruskalmetric} with \eqref{eq:kruskalinversion} 
can be smoothly extended from Region~I, where $(U,V) \in \BbbR^- \times \BbbR^+$, to $(U,V) \in \BbbR \times \BbbR$, as summarised in Table \ref{tab:exterior} and illustrated in Figure~\ref{RNConformal}: the extension covers Regions~I, II, II' and I' as shown in Figure~\ref{RNConformal}, and the boundaries at which they are joined. 
We call this spacetime~$\mathcal{M}_K$. 
In Regions II and~II', where $UV>0$, $r \in (r_-, r_+)$ 
is determined as a function of $U$ and $V$ from 
\begin{align}
UV = e^{2\kappa_+ {\tilde r}_*} , 
\label{eq:kruskalinversion-interiors}
\end{align}
where the relation between $r$ and ${\tilde r}_*$ is determined by
\begin{align}
d {\tilde r}_* = \frac{dr}{F(r)} ,
\label{eq:tortoise-interior-def}
\end{align} 
with the additive constant in ${\tilde r}_*$ chosen so that the extension of the metric function 
$F(r)\left( \kappa_+^2 U V \right)^{-1}$ in \eqref{eq:kruskalmetric} across $UV=0$ is smooth. 
If $F$ is real analytic, the extended metric is real analytic. 
Note that $\xi$ extends smoothly from Region I to~$\mathcal{M}_K$, 
having the formula $\xi = \kappa_+ (-U\partial_U + V\partial_V)$, 
and $\xi$ has a bifurcate Killing horizon at $UV=0$, where $r=r_+$. 

The coordinates $(U,V)$ do not extend to $r=r_-$. 
Another set of Kruskal-type coordinates can be introduced 
to cover the four regions joined at the Killing horizon $r=r_-$; 
these coordinates will however not be needed for what follows.

\subsubsection{Hybrid coordinates}

Consider the region where $-\infty < U < \infty$ and $0 < V < \infty$. 
In Figure~\ref{RNConformal}, this consists of Regions I and II and their joint boundary, the black hole horizon~$H^F$. 
We call this spacetime~$\mathcal{M}_U$. 

Given the Kruskal-like coordinates $(U,V) \in \BbbR \times \BbbR^+$ in $\mathcal{M}_U$, 
we introduce the new coordinates $(U,v)\in \BbbR \times \BbbR$ in $\mathcal{M}_U$ 
by~\eqref{eq:EF-to-Kruskal-V}. 
We refer to these as the hybrid coordinates, 
being Kruskal-like in $U$ and Eddington-Finkelstein-like in~$v$. 
The metric takes the form 
\begin{equation}
ds^2 = \frac{F(r)}{\kappa_+ U} \, dU \, dv, 
\label{eq:half-and-half-metric}
\end{equation}
where $r$ is determined as a function of $U$ and $v$ from 
\begin{subequations}
\label{eq:half-and-half-inversion}
\begin{align}
- U e^{\kappa_+ v} & = e^{2\kappa_+ r_*}
\ \ \text{for} \ \ U<0, 
\\
U e^{\kappa_+ v} & = e^{2\kappa_+ {\tilde r}_*}
\ \ \text{for} \ \ U>0, 
\end{align}
\end{subequations}
with $r$ related to $r_*$ and ${\tilde r}_*$ as above. 
The black hole horizon $H^F$ is at $U=0$.

\subsubsection{Eddington-Finkelstein coordinates}

For completeness, we record here how the standard ingoing Eddington-Finkelstein 
coordinates are related to the coordinate systems introduced above. 

In $\mathcal{M}_U$, 
starting from \eqref{eq:half-and-half-metric} and replacing $U$ by $r$ 
puts the metric in the Eddington-Finkelstein form 
\begin{equation}
ds^2 = - F(r) dv^2 + 2 dv \, dr , 
\label{eq:edd-fink-metric}
\end{equation}
where $r_- < r < \infty$ and $v\in\BbbR$. 
This metric can be extended to $0< r < \infty$, covering also 
$H_-^F$ and 
Region III in Figure~\ref{RNConformal}. 
The usual way to obtain \eqref{eq:edd-fink-metric}
is to start from Region I with the metric~\eqref{eq:RNcoords-metric}, 
set $dt = dv - dr/F(r)$, and then allow $0< r < \infty$.

\subsection{Timelike geodesics approaching the Cauchy horizon\label{subsec:infalling-geodesic}}

We are interested in timelike geodesics in the black hole interior, 
Region II in Figure~\ref{RNConformal}. 

It is convenient to introduce in Region II 
the interior Schwarzschild-like coordinates 
$(\tilde{t},r)$, 
in which the metric reads 
\begin{align}
ds^2 = \frac{dr^2}{F(r)} -F(r) d{\tilde{t}\,}^2 , 
\label{eq:RNcoords-interior-metric}
\end{align}
where now $F(r)<0$, $r \in (r_-, r_+)$ is a timelike coordinate decreasing to the future, 
and $\tilde{t} \in \BbbR$ is a spacelike coordinate increasing to the right. 
These coordinates may be obtained from 
\eqref{eq:edd-fink-metric} by writing $dv = d\tilde{t} + dr/F(r)$, 
or from \eqref{eq:kruskalmetric} by writing first 
\begin{subequations}
\label{eq:int-kruskal-to-eddfink}
\begin{align}
U &= e^{\kappa_+ \tilde{u}}, 
\\
V &= e^{\kappa_+ \tilde{v}}, 
\end{align}
\end{subequations}
where $(U,V) \in \BbbR_+ \times \BbbR_+$ 
and $({\tilde u},{\tilde v}) \in \BbbR \times \BbbR$, 
and then
\begin{subequations}
\label{eq:int-eddfink-to-schw}
\begin{align}
\tilde{u} &= {\tilde r}_* - \tilde{t} , 
\\
\tilde{v} &= {\tilde r}_* + \tilde{t} , 
\end{align}
\end{subequations}
and finally using \eqref{eq:tortoise-interior-def} 
to replace ${\tilde r}_*$ by~$r$. 

From \eqref{eq:RNcoords-interior-metric}, 
it is now straightforward to verify that 
the timelike geodesics are the integral curves of the system 
\begin{subequations}
\label{eqmotion-interior}
\begin{align}
\dot{\tilde{t}} &= \frac{E}{F(r)}, 
\label{tildet-eq}
\\
\dot{r} &= - \sqrt{E^2-F(r)}, 
\label{eq:rdot-interior}
\end{align}
\end{subequations}
where the overdot denotes derivative with respect to the proper time, increasing to the future, 
and $E\in \BbbR$ is a constant of integration. 
In the coordinates $({\tilde u},{\tilde v})$, the system \eqref{eqmotion-interior}
reads 
\begin{subequations}
\label{eq:tildedots}
\begin{align}
\dot {\tilde u} &= \frac{1}{\sqrt{E^2 - F(r)} - E} , 
\label{eq:dot-utilde}
\\
\dot {\tilde v} &= \frac{1}{\sqrt{E^2 - F(r)} + E} . 
\end{align}
\end{subequations}

All these geodesics hit the Cauchy horizon at $r=r_-$ in finite proper time. 
A~geodesic with $E>0$ 
travels towards decreasing~$\tilde{t}$, 
crossing the Cauchy horizon's left branch $H^L_-$ into Region~III, 
as is perhaps most easily seen in the Eddington-Finkelstein 
coordinates~\eqref{eq:edd-fink-metric}; 
similarly, a geodesic with $E<0$ 
travels towards increasing~$\tilde{t}$, 
crossing the Cauchy horizon's right branch $H^R_-$ into Region~III'. 
A geodesic with $E=0$ crosses the bifurcation point where $H^L_-$ and $H^R_-$ meet, 
entering a new region in which $r \in (r_-, r_+)$. 

We note that a geodesic with $E>0$ continues in the past to Region~I, having fallen in from there, 
a geodesic with $E<0$ has fallen in from Region~I', and a geodesic with $E=0$ 
has emerged from Region~II', the white hole, through the bifurcation 
point where Regions I, I', II and II' meet. 
We shall however not consider these geodesics beyond Region~II, in the past or in the future.

\section{Quantum scalar field\label{seq:quantum-scalar}}

Let $\phi$ be a real massless scalar field, 
with the field equation 
\begin{align}
\Box \phi = 0 . 
\end{align}

In $\mathcal{M}_K$, it follows from the conformal invariance of the massless scalar field, 
and the conformally flat form of the metric given in~\eqref{eq:kruskalmetric}, 
that $\phi$ has a Fock quantisation based on the input encoded in the Kruskal coordinates. 
The field equation reads 
\begin{align}
\partial_U \partial_V \phi =0 , 
\end{align}
and a Fock quantisation is obtained by defining positive frequencies in terms of $\partial_U$ and~$\partial_V$. 
The corresponding vacuum state is known as the 
Hartle-Hawking-Israel (HHI) state $| 0_\text{H} \rangle$\cite{Hartle:1976tp,Israel:1976ur}. 
The Wightman function in $| 0_\text{H} \rangle$ is given by 
\begin{align}
\mathcal{W}_\text{H}(\mathsf{x},\mathsf{x'}) 
& \doteq 
\langle 0_\text{H} | \phi(\mathsf{x}) \phi(\mathsf{x'}) | 0_\text{H} \rangle 
\notag\\
&= -\frac{1}{4 \pi} \ln \left[(\epsilon+i \Delta U) (\epsilon + i \Delta V) \right] , 
\label{eq:HHI-wightman}
\end{align}
where we have written 
$\mathsf{x} = (U,V)$ and $\mathsf{x}' = (U',V')$, with $\Delta U = U-U'$ and $\Delta V = V-V'$. 
The logarithm denotes the branch that is real-valued for positive argument, 
and the limit $\epsilon \to 0_+$ is understood. 

In $\mathcal{M}_U$, it follows from the conformally flat form 
of the metric given in~\eqref{eq:half-and-half-metric} 
that $\phi$ has a Fock quantisation based on the input encoded in the hybrid coordinates, 
and this quantisation is inequivalent to that obtained by restriction of the above Fock quantisation in $\mathcal{M}_K$. 
The field equation reads 
\begin{align}
\partial_U \partial_v \phi =0 , 
\end{align}
and a Fock quantisation is obtained by defining positive frequencies in terms of $\partial_U$ and~$\partial_v$. 
The corresponding vacuum state is known as the Unruh state $| 0_\text{U} \rangle$\cite{Unruh:1976db}. 
The Wightman function in $| 0_\text{U} \rangle$ is given by 
\begin{align}
\mathcal{W}_\text{U}(\mathsf{x},\mathsf{x'}) 
& \doteq 
\langle 0_\text{U} | \phi(\mathsf{x}) \phi(\mathsf{x'}) | 0_\text{U} \rangle 
\notag\\
&= -\frac{1}{4 \pi} \ln \left[(\epsilon+i \Delta U) (\epsilon + i \Delta v) \right] , 
\label{eq:U-wightman}
\end{align}
where the notation is as in \eqref{eq:HHI-wightman} but now with $\Delta v = v-v'$. 
The logarithm denotes again the branch that is real-valued for positive argument, 
and the limit $\epsilon \to 0_+$ is understood. 

$| 0_\text{H} \rangle$ is by construction regular in $\mathcal{M}_K$ and 
$| 0_\text{U} \rangle$ is regular in~$\mathcal{M}_U$, 
in the sense that the short distance behaviour of both 
$\mathcal{W}_\text{H}$ and $\mathcal{W}_\text{U}$ 
satisfies the Hadamard condition~\cite{Decanini:2005eg}. 
Observers at constant $r$ in Region I experience 
$| 0_\text{H} \rangle$ as a thermal equilibrium state, 
in the local Hawking temperature 
$T_H(r) = \kappa_+/\bigl(2\pi \sqrt{F(r)}\,\bigr)$\cite{Hartle:1976tp,Israel:1976ur}, 
whereas these observers experience $| 0_\text{U} \rangle$ 
as a state in which the ingoing part of the field is in a vacuum-like state but the outgoing part is in the local Hawking temperature~$T_H$\cite{Unruh:1976db}. 
$| 0_\text{U} \rangle$~mimics the late-time properties of a state that ensues 
from the collapse of an initially static star\cite{Unruh:1976db,Davies:1976ei,Juarez-Aubry:2018ofz}. 

Both $\mathcal{W}_\text{H}$ and $\mathcal{W}_\text{U}$ have an infrared ambiguity, 
characteristic of a massless field in $1+1$ dimensions, 
and we have resolved this ambiguity as shown in 
\eqref{eq:HHI-wightman} and~\eqref{eq:U-wightman}. 
$\mathcal{W}_\text{H}$~is invariant under the isometry generated by~$\xi$. 
$\mathcal{W}_\text{U}$~is invariant under this isometry only up
to an additive constant; further, our formula \eqref{eq:U-wightman} for $\mathcal{W}_\text{U}$ 
contains dimensionally inconsistent notation in that $U$ is dimensionless but $v$ has the dimension of length. However, in the rest of the paper we shall probe 
$\mathcal{W}_\text{H}$ and $\mathcal{W}_\text{U}$ by means that involve derivatives: 
under this probing both $\mathcal{W}_\text{H}$ and $\mathcal{W}_\text{U}$ 
will give answers that are invariant under the isometry generated by~$\xi$, 
and the dimensional inconsistency of \eqref{eq:U-wightman} will drop out. 
A similar issue in the $(1+1)$-dimensional Schwarzschild spacetime was 
discussed in~\cite{Juarez-Aubry:2014jba,Juarez-Aubry:2018ofz}.

\section{Unruh-DeWitt detector: transition probability and transition rate\label{sec:detector}} 

In this section we first briefly review the technical and conceptual aspects 
of probing a quantum field with time-and-space localised quantum systems known as 
Unruh-DeWitt (UDW) detectors\cite{Unruh:1976db,DeWitt:1979}. 
We then specialise to a spatially pointlike detector, 
coupled linearly to the proper time derivative of the scalar field, 
and treated to first order in perturbation theory. 

\subsection{A quantum detector localised in time and space}

We wish to probe the quantum field with a time-and-space localised quantum system known 
as an Unruh-DeWitt (UDW) detector\cite{Unruh:1976db,DeWitt:1979}: 
a quantum system that moves through the spacetime on the timelike worldline~$\mathsf{x}(\tau)$, 
parametrised by the proper time~$\tau$. 
What needs to be specified 
is the detector's internal dynamics, 
the sense of localisation, 
and the coupling. 

For the internal dynamics, we assume that the detector is a two-level system. 
The Hilbert space is spanned by two orthonormal states, 
with the respective eigenenergies $0$ and $\omega \in \BbbR \setminus \{0\}$, 
defined with respect to~$\tau$. 
For $\omega>0$, the state with eigenenergy $0$ is the ground state and the state with eigenenergy 
$\omega$ is the excited state; for $\omega<0$, the roles of the states are reversed. 

Generalisations to detectors with multiple levels could be considered. 
For example, a detector that has the dynamics of a harmonic oscillator 
is convenient when the coupling between the field and the 
detector is analysed nonperturbatively\cite{Lin:2006jw,Lin:2015aua}. 
Multiple-level systems however reduce to two-level systems when 
treated in first-order perturbation theory,
and this is what we shall do below. 

For the localisation in space, 
we assume that the detector's spatial size is negligible, 
as in the detector model introduced by DeWitt\cite{DeWitt:1979}: 
the detector is restricted strictly to the worldline~$\mathsf{x}(\tau)$. 
This will make the coupling between the field and the detector slightly singular, 
but the singularity will not produce infinities in the first-order perturbative 
treatment that we shall follow below. 
Allowing the detector to have a nonzero spatial size, 
as in the detector model originally introduced by Unruh\cite{Unruh:1976db}, 
would present a technical challenge for formulating the notion of a spatial profile when the spacetime is curved, or even in flat spacetime when the detector's motion is non-inertial\cite{Schlicht:2003iy,Louko:2006zv,Langlois:2005if,DeBievre:2006pys,Kolekar:2019ljv}; 
further, a finite spatial size would raise questions about the relativistic consistency of the coupled system, and about the sense in which the two-level detector approximates an underlying more fundamental detection described by quantum fields\cite{Bostelmann:2020unl,Martin-Martinez:2020lul,Ruep:2021fjh,Grimmer:2021qib}. 

For the localisation in time, we assume that the detector operates for a finite interval of proper time. 
As we wish to consider a strongly time-dependent situation, we shall consider the limit in which the switch-on and switch-off are instantaneous. 
While this limit creates a divergence in the detector's transition probability, 
the divergence is a pure switching effect, 
and the time-dependent features can be extracted by considering the transition rate, rather than the transition probability, as we shall discuss below in Section~\ref{sec:transrate}. 

For the coupling between the field and the detector, 
a frequently-considered choice is to couple the detector linearly to~$\phi\bigl(\mathsf{x}(\tau)\bigr)$, that is, to the value of the field $\phi$ at the location of the detector: this model is known to capture the essential features of light-matter interaction when angular momentum interchange is negligible\cite{Martin-Martinez:2012ysv,Alhambra:2013uja}. In our case of a massless field in $1+1$ spacetime dimensions, 
this choice however inherits the infrared ambiguity of the Wightman function. 
We therefore couple the detector linearly to~$\partial_\tau\phi\bigl(\mathsf{x}(\tau)\bigr)$, 
that is, to the proper time derivative of $\phi$ at the location of the detector, 
which will cure the infrared ambiguity. 
A~selection of previous work on a derivative-coupled detector in a range of contexts is
available in\cite{Juarez-Aubry:2014jba,Raval:1995mb,Raine:1991kc,Wang:2013lex,Juarez-Aubry:2018ofz,Davies:2002bg,Martin-Martinez:2014qda,Louko:2014aba,Brown:2015yma,Thinh:2016fom,Moustos:2018get,Louko:2018pij,Cong:2018vqx,Tjoa:2020riy,Tjoa:2020eqh,Kollas:2021nqy}. 

Nonlinear couplings could be considered, 
but they would typically require additional regularisation\cite{Hummer:2015xaa}. 
We shall consider the linear coupling to $\partial_\tau\phi\bigl(\mathsf{x}(\tau)\bigr)$.

\subsection{Spatially pointlike detector with a linear derivative coupling\label{sec:transrate}}

To recap, we consider a spatially pointlike two-level detector, on the timelike worldline~$\mathsf{x}(\tau)$, 
parametrised by the proper time~$\tau$, 
coupled linearly to $\partial_\tau\phi\bigl(\mathsf{x}(\tau)\bigr)$. 

Working to first-order perturbation theory in the coupling between the detector and the field, 
the probability of the detector to make a transition from 
the eigenenergy $0$ state to the eigenenergy $\omega$ state 
is a multiple of the response function $\mathcal{F}(\omega)$, given by 
\begin{align}
\mathcal{F}(\omega)&= \int d\tau' \, d\tau'' 
\, \chi(\tau') \chi(\tau'') 
\, e^{-i\omega (\tau'-\tau'')} \, 
\notag\\
& \hspace{8ex}
\times \partial_{\tau'} \partial_{\tau''} \mathcal{W}(\tau',\tau''), 
\label{eq:responsefunction-formula}
\end{align}
where $\mathcal{W}(\tau',\tau'') = \langle\Psi|\phi\bigl(\mathsf{x}(\tau')\bigr)\phi\bigl(\mathsf{x}(\tau'')\bigr)|\Psi\rangle$ 
is the pullback of the scalar field's Wightman function to the detector's worldline, 
$|\Psi\rangle$ denotes the initial state of the field, and 
the real-valued switching function $\chi$ specifies how the interaction is turned on and off. 
When $|\Psi\rangle$ is a state satisfying the Hadamard short-distance condition\cite{Decanini:2005eg}, 
$\mathcal{W}(\tau',\tau'')$ is a well-defined distribution under 
mild assumptions about the detector's trajectory~\cite{hormander-book,Fewster:1999gj}, 
and $\mathcal{F}(\omega)$ is well defined under mild assumptions about~$\chi$; 
for example, taking $\chi$ to be smooth and of compact support suffices. 
As the factor relating $\mathcal{F}(\omega)$ to the probability depends only on the detector's internal structure, we 
refer to $\mathcal{F}(\omega)$ as the transition probability, with a minor abuse of terminology. 
Note that the derivatives in \eqref{eq:responsefunction-formula} are responsible for making 
the infrared ambiguity of $\mathcal{W}$ drop out of~$\mathcal{F}$.  

The response function $\mathcal{F}(\omega)$ \eqref{eq:responsefunction-formula} 
depends not just on the quantum field's initial state $|\Psi\rangle$ and the detector's trajectory, 
but also on the switching function~$\chi$. 
To consider the response of the detector as it approaches the Cauchy horizon, 
we consider a $\chi$ that cuts off at a sharply-defined moment of proper time, 
shortly before the trajectory reaches the horizon. 
This creates a technical issue: if the detector is switched on sharply at proper time $\tau_0$ and off at proper time $\tau>\tau_0$, so that $\chi(u) = \Theta(\tau-u)\Theta(u-\tau_0)$, $\mathcal{F}(\omega)$ becomes divergent, due to the large contributions from the switch-on and switch-off moments; 
the issue for a derivative-coupling detector in $1+1$ dimensions 
is the same as for a non-derivative-coupling detector in $3+1$ dimensions \cite{Schlicht:2003iy,Louko:2006zv,Satz:2006kb,Obadia:2007qf,Louko:2007mu,Juarez-Aubry:2014jba}. 
To circumvent this issue, we shall not consider the sharp switching limit of the transition probability $\mathcal{F}(\omega)$, 
but we consider instead the transition rate, 
the derivative of this probability with respect to the switch-off moment, 
which has a finite limit when the switching becomes sharp. 
Denoting the transition rate by $\dot{\mathcal{F}}(\omega,\tau, \tau_0)$, where $\tau_0$ and $\tau$ are respectively the switch-on and switch-off proper times, 
we have\cite{Juarez-Aubry:2014jba}
\begin{align}
\dot{\mathcal{F}}(\omega,\tau, \tau_0)  & = -\omega \Theta(-\omega) 
\notag\\
& \hspace{3ex}
+ \frac{1}{\pi} \! \left(\frac{\cos(\omega\Delta \tau)}{\Delta \tau} + |\omega| \si(|\omega|\Delta \tau) \right) 
\notag\\
& \hspace{3ex}
+ 2 \int_{\tau_0}^{\tau} \! d\tau' \, \text{Re} \Biggl[ e^{-i \omega (\tau-\tau')} 
\Biggl( \partial_{\tau} \partial_{\tau'} \mathcal{W}(\tau,\tau')
\notag\\
& \hspace{19ex}
 + \frac{1}{2 \pi {(\tau-\tau')}^2} \Biggr) \Biggr],
\label{Fdot}
\end{align}
where $\Delta \tau \doteq \tau - \tau_0$ and $\si$ is the sine integral function\cite{NIST}. 
Note that the integrand in \eqref{Fdot} is nonsingular at $\tau'\to\tau$ 
because of the Hadamard property of the Wightman function\cite{Decanini:2005eg}. 

We shall use \eqref{Fdot} to 
address the behaviour of $\dot{\mathcal{F}}(\omega,\tau, \tau_0)$ 
near the Cauchy horizon in Section~\ref{sec:detector-near}.

\section{Detector near the Cauchy horizon\label{sec:detector-near}} 

We now specialise to a detector on a geodesic in Region~II, 
as described in Section~\ref{subsec:infalling-geodesic}, 
and we specialise to the HHI and Unruh states as described in Section~\ref{seq:quantum-scalar}. 
We shall find the leading behaviour of the transition rate as the detector approaches the Cauchy horizon. 

Let $\tau_h$ be the value of the proper time at which the trajectory hits the Cauchy horizon. 
In the notation of~\eqref{Fdot}, we then have $\tau_0 < \tau < \tau_h$, 
where $\tau_0$ is the switch-on moment and $\tau$ is the switch-off moment. 
For concreteness, we assume that the switch-on moment $\tau_0$ is in Region~II, 
for all trajectories and all states. 
We consider the asymptotic behaviour of 
$\dot{\mathcal{F}}(\omega,\tau, \tau_0)$ \eqref{Fdot} as $\tau\to\tau_h$, 
with $\tau_0$ fixed. 

Note first that the terms outside the integral in \eqref{Fdot} are of order ${\cal O}(1)$ as $\tau\to\tau_h$. 
Also, note that by \eqref{eq:HHI-wightman} and~\eqref{eq:U-wightman}, the imaginary part of $\mathcal{W}(\tau,\tau')$ is a constant for $\tau' < \tau$. We hence have 
\begin{align}
& \dot{\mathcal{F}}(\omega,\tau, \tau_0)  = 
2 \int_{\tau_0}^{\tau} \! d\tau' \, \cos\bigl(\omega (\tau-\tau')\bigr) 
\notag\\
& \hspace{3ex}
\times 
\left( \partial_{\tau} \partial_{\tau'} \mathcal{W}(\tau,\tau') + \frac{1}{2 \pi {(\tau-\tau')}^2} \right) 
\ \ + {\mathcal O}(1) . 
\end{align}
Integrating by parts gives 
\begin{align}
& \dot{\mathcal{F}}(\omega, \tau, \tau_0) = - 2 \cos\left(\omega \Delta \tau\right)\partial_\tau \mathcal{W}(\tau,\tau_0)
\nonumber \\
&
+ 2 \lim_{\tau' \rightarrow \tau } \! \left(\partial_\tau \mathcal{W}(\tau,\tau')+\frac{1}{2 \pi (\tau-\tau')} \right) \nonumber \\
&
- 2 \omega \int_{\tau_0}^\tau \! d \tau' \sin \bigl(\omega(\tau-\tau')\bigr) \! \left(\partial_\tau \mathcal{W}(\tau,\tau')+\frac{1}{2 \pi (\tau-\tau')} \right) 
\notag\\
& + \mathcal{O}(1),
\label{eq:genstatelim}
\end{align}
where the second term on the right-hand side is well defined and finite by the Hadamard property of the state\cite{Decanini:2005eg}. 

What remains is to estimate \eqref{eq:genstatelim} as $\tau\to\tau_h$, 
for the HHI and Unruh states, and for a detector approaching the Cauchy horizon 
on the left branch~$H^L_-$, 
on the right branch~$H^R_-$, and at the bifurcation point. 
We address the three different parts of the Cauchy horizon 
in respectively Appendix \ref{app:HLminus-asympt}, \ref{app:HRminus-asympt} and~\ref{app:Hbif}. 
We collect the outcomes here. 

In the HHI state, we find
\begin{subequations}
\label{eq:fdot-final-hhi}
\begin{align}
\dot{\mathcal{F}}_{\text{H}}^{L}(\omega, \tau, \tau_0) 
&= \dot{\mathcal{F}}_{\text{H}}^{R}(\omega, \tau, \tau_0) 
\notag\\
&= \frac{1}{4 \pi (\tau_h-\tau)} \! \left(\frac{\kappa_+}{\kappa_-} - 1 
+o(1) \right) , 
\label{eq:fdot-final-hhi-LR}
\\
\dot{\mathcal{F}}_{\text{H}}^{0}(\omega, \tau, \tau_0) 
&= \frac{1}{2 \pi (\tau_h-\tau)} \! \left(\frac{\kappa_+}{\kappa_-} - 1 
+o(1) \right) , 
\label{eq:fdot-final-hhi-0}
\end{align}
\end{subequations}
where the superscript $L$, $R$, $0$ indicates respectively 
$H^L_-$, $H^R_-$ and the bifurcation point. 
When $\kappa_+ \ne \kappa_-$, the leading term hence diverges proportionally to 
$1/(\tau_h-\tau)$, in all three cases, 
with an overall sign that differs for $\kappa_- < \kappa_+$ and $\kappa_+ < \kappa_-$. 
That the responses on approaching $H^L_-$ and $H^R_-$ are identical follows 
from the left-right symmetry of the HHI state, and the divergence for $H^L_-$ (respectively $H^R_-$) 
comes only from the right-moving (respectively left-moving) part of the field. 
The divergence on approaching the bifurcation point gets contributions from both parts of the field, leading to the double strength in~\eqref{eq:fdot-final-hhi-0}. 

In the special case $\kappa_- = \kappa_+$, 
the leading term in \eqref{eq:fdot-final-hhi} vanishes. 
In this case the error terms in \eqref{eq:fdot-final-hhi} can be tightened, 
as shown in appendices \ref{app:HLminus-asympt}, \ref{app:HRminus-asympt} and~\ref{app:Hbif}, 
with the outcome that the transition rate remains bounded on approaching the horizon. 

In the Unruh state, we find
\begin{subequations}
\label{eq:fdot-final-u}
\begin{align}
\dot{\mathcal{F}}_{\text{U}}^{L}(\omega, \tau, \tau_0) 
&= \frac{1}{4 \pi (\tau_h-\tau)} \! \left(\frac{\kappa_+}{\kappa_-} - 1 
+o(1) \right) , 
\label{eq:fdot-final-u-left}
\\
\dot{\mathcal{F}}_{\text{U}}^{R}(\omega, \tau, \tau_0) 
&= - \frac{1 + o(1)}{4 \pi (\tau_h-\tau)} , 
\label{eq:fdot-final-u-right}
\\
\dot{\mathcal{F}}_{\text{U}}^{0}(\omega, \tau, \tau_0) 
&= \frac{1}{4 \pi (\tau_h-\tau)} \! \left(\frac{\kappa_+}{\kappa_-} - 2 
+o(1) \right) , 
\label{eq:fdot-final-u-bif}
\end{align}
\end{subequations}
The divergence on approaching $H^L_-$ is as in the HHI state, 
but the divergence on approaching $H^R_-$ is independent of the surface gravities, 
and has always a negative sign. The divergence on approaching the bifurcation point is the sum. 


In the special case $\kappa_- = \kappa_+$, the leading term in \eqref{eq:fdot-final-u-left} vanishes, and we show in Appendix \ref{app:HLminus-asympt} that $\dot{\mathcal{F}}_{\text{U}}^{L}$ remains bounded on approaching~$H^L_-$. In the special case $\kappa_+ = 2\kappa_-$, 
the leading term in \eqref{eq:fdot-final-u-bif} vanishes, and we show in Appendices \ref{app:HRminus-asympt} and \ref{app:Hbif} that 
\begin{align}
\dot{\mathcal{F}}_{\text{U}}^{0}(\omega, \tau, \tau_0) 
&= \frac{1 + o(1)}{2\pi (\tau_h-\tau)\bigl(- \ln(\tau_h-\tau) \bigr)} ,  
\label{eq:fdot-final-u-bif-special}
\end{align}
which diverges on approaching the bifurcation point, but less quickly than $1/(\tau_h-\tau)$.

\section{Energy near the Cauchy horizon\label{sec:stress-energy}}

In this section we compare the 
the transition rate results of Section \ref{sec:detector-near}
to the energy density seen by a geodesic observer.

\subsection{HHI state}

Recall from \eqref{eq:kruskalmetric} that in the Kruskal coordinates $(U,V)$ we have 
\begin{equation}
ds^2 = \Omega_K^2 (- \, dU \, dV) , 
\end{equation}
where 
\begin{equation}
\Omega_K^2 = - \frac{F}{\kappa_+^2 U V} , 
\end{equation}
and $r$ is determined as a function of $r$ by \eqref{eq:kruskalinversion} for $UV<0$ and by \eqref{eq:kruskalinversion-interiors} for $UV>0$. 
As the HHI state is built on the positive frequency definition 
provided by $\partial_U$ and~$\partial_V$, conformal invariance of the field shows that the renormalised stress-energy tensor $T_{ab}^{\text{H}}$ 
in the HHI state is given by \cite{Davies:1976ei,Wald:1978ce,Birrell:1982ix}
\begin{align}
T_{ab}^{\text{H}}(\mathsf{x})
= \Xi_{ab}(\mathsf{x})
-\frac{R(\mathsf{x})}{48 \pi} g_{ab}, 
\label{Tren-HHI}
\end{align}
where 
\begin{subequations}
\begin{align}
\Xi_{UU} &= -(1/12\pi) \Omega_K \partial_U^2 \Omega_K^{-1}, \\
\Xi_{VV} &= -(1/12\pi) \Omega_K \partial_V^2 \Omega_K^{-1}, \\
\Xi_{UV} &=  \Xi_{VU} = 0, 
\end{align}
\end{subequations}
and $R$ is the Ricci scalar. 

The energy density on a timelike worldline parametrised by the proper time 
$\tau$ is given by  
\begin{align}
\mathcal{E}_{\text{H}}
&= \dot{x}^a \dot{x}^b
T_{ab}^{\text{H}}
\notag\\
& = \dot{U}^2 \Xi_{UU}
+ \dot{V}^2 \Xi_{VV}
+ \frac{R}{48 \pi}
\notag\\
&= 
- \frac{\left( \dot{\tilde{u}}^2 + \dot{\tilde{v}}^2\right)}{192 \pi} 
\left( {F'}^2 - 2 F F'' - 4\kappa_+^2 \right)
+ \frac{R}{48 \pi} , 
\end{align}
where the overdots denote derivative with respect to~$\tau$, 
and the last expression, 
using the coordinates $(\tilde{u}, \tilde{v})$, 
assumes the worldline to be in Region~II\null. Near the Cauchy horizon, we have 
\begin{align}
{F'}^2 - 2 F F''
= 
4\kappa_-^2 
+ \mathcal{O} \bigl((r-r_-)^2\bigr) , 
\end{align}
and the Ricci scalar remains bounded. For a geodesic approaching the Cauchy horizon, 
it then follows from the estimates given for $\dot{\tilde{u}}(\tau)$, $\dot{\tilde{v}}(\tau)$ and $r(\tau)$ in Appendices 
\ref{app:HLminus-asympt}, 
\ref{app:HRminus-asympt}
and 
\ref{app:Hbif}
that 
\begin{subequations}
\label{eq:Edens-final-hhi}
\begin{align}
\mathcal{E}_{\text{H}}^L(\tau) 
&= \mathcal{E}_{\text{H}}^R(\tau)  
\notag\\
&= \frac{\bigl(1 + \mathcal{O}(\tau_h-\tau) \bigr)}{48 \pi {(\tau_h-\tau)}^2} \! \left(\frac{\kappa_+^2}{\kappa_-^2} - 1 
\right) + \mathcal{O}(1) , 
\\
\mathcal{E}_{\text{H}}^0(\tau) 
&= \frac{1}{24 \pi {(\tau_h-\tau)}^2} \! \left(\frac{\kappa_+^2}{\kappa_-^2} - 1 
\right)  + \mathcal{O}(1) , 
\end{align}
\end{subequations}
where the superscript $L$, $R$ and $0$ 
indicates whether the geodesic approaches the Cauchy horizon at $H^L_-$, $H^R_-$ 
or the bifurcation point. 
Note that when $\kappa_+ = \kappa_-$, 
the energy density remains bounded in all three cases. 

\subsection{Unruh state}

We proceed similarly with the Unruh state. In the hybrid coordinates $(U,v)$, 
the metric \eqref{eq:half-and-half-metric} reads 
\begin{equation}
ds^2 = \Omega_U^2 (- \, dU \, dv) , 
\end{equation}
where 
\begin{equation}
\Omega_U^2 = - \frac{F}{\kappa_+ U} , 
\end{equation}
and $r$ is determined as a function of $U$ and $v$ from~\eqref{eq:half-and-half-inversion}. 
The renormalised stress-energy tensor $T_{ab}^{\text{U}}$ 
in the Unruh state is given by 
\begin{align}
T_{ab}^{\text{U}}(\mathsf{x})
= \Psi_{ab}(\mathsf{x})
-\frac{R(\mathsf{x})}{48 \pi} g_{ab}, 
\label{Tren-Unruh}
\end{align}
where 
\begin{subequations}
\begin{align}
\Psi_{UU} &= -(1/12\pi) \Omega_U \partial_U^2 \Omega_U^{-1}, \\
\Psi_{vv} &= -(1/12\pi) \Omega_U \partial_v^2 \Omega_U^{-1}, \\
\Psi_{Uv} &=  \Psi_{vU} = 0 . 
\end{align}
\end{subequations}

The energy density on a timelike worldline parametrised by the proper time 
$\tau$ is now given by 
\begin{align}
\mathcal{E}_{\text{U}}
&= \dot{x}^a \dot{x}^b
T_{ab}^{\text{U}}
\notag\\
& = \dot{U}^2 \Psi_{UU}
+ \dot{v}^2 \Psi_{vv}
+ \frac{R}{48 \pi}
\notag\\
&= 
- \frac{\dot{\tilde{u}}^2 }{192 \pi} 
\left( {F'}^2 - 2 F F'' - 4\kappa_+^2 \right)
\notag\\
& \hspace{3ex}
- \frac{\dot{v}^2 }{192 \pi} 
\left( {F'}^2 - 2 F F'' \right)
+ \frac{R}{48 \pi} , 
\end{align}
where the last expression assumes the worldline to be in Region~II\null. 
For a geodesic approaching the Cauchy horizon, 
proceeding as with \eqref{eq:Edens-final-hhi} and using the same notation, 
we find 
\begin{subequations}
\label{eq:Edens-final-unruh}
\begin{align}
\mathcal{E}_{\text{U}}^L(\tau) 
&= \frac{\bigl(1 + \mathcal{O}(\tau_h-\tau) \bigr)}{48 \pi {(\tau_h-\tau)}^2} \! \left(\frac{\kappa_+^2}{\kappa_-^2} - 1 
\right) + \mathcal{O}(1) , 
\\
\mathcal{E}_{\text{U}}^R(\tau) 
&= - \frac{1 + \mathcal{O}(\tau_h-\tau)}{48 \pi {(\tau_h-\tau)}^2} , 
\\
\mathcal{E}_{\text{U}}^0(\tau) 
&= \frac{1}{48 \pi {(\tau_h-\tau)}^2} \! \left(\frac{\kappa_+^2}{\kappa_-^2} - 2 
\right)  + \mathcal{O}(1) . 
\end{align}
\end{subequations}
Note that $\mathcal{E}_{\text{U}}^L$ remains bounded when $\kappa_+ = \kappa_-$, 
$\mathcal{E}_{\text{U}}^0$ remains bounded when $\kappa_+ = \sqrt{2} \, \kappa_-$, 
and $\mathcal{E}_{\text{U}}^R$ diverges for all values of $\kappa_+$ and~$\kappa_-$.

\subsection{Comparison}

Comparing \eqref{eq:fdot-final-hhi} with~\eqref{eq:Edens-final-hhi}, and 
\eqref{eq:fdot-final-u} with~\eqref{eq:Edens-final-unruh}, 
we see that there is a significant qualitative and quantitative agreement between 
the divergence of the detector's transition rate and the divergence of the observer's energy density on approaching the Cauchy horizon.  

For both the HHI and Unruh states, neither quantity diverges on $H_-^L$ when $\kappa_+ = \kappa_-$, whereas both quantities diverge for $\kappa_+ \ne \kappa_-$, and the sign of the divergence agrees, being positive for $\kappa_- < \kappa_+$ and negative for  $\kappa_+ < \kappa_-$. 
On $H_-^R$, the situation for the HHI state is similar, whereas for the Unruh state there is always a divergence with a negative overall coefficient. 
The Cauchy horizon bifurcation point interpolates between the two branches; in the Unruh state, 
the threshold between positive and negative divergence occurs at $\kappa_+ = 2 \kappa_-$ with the transition rate and at $\kappa_+ = \sqrt{2} \, \kappa_-$ with the energy density. 

When a divergence occurs, it is proportional to $(\tau_h - \tau)^{-1}$ in the transition rate and proportional to $(\tau_h - \tau)^{-2}$ in the energy density. 
In the integral of the energy density over a finite proper time interval, 
the divergence is proportional to $(\tau_h - \tau)^{-1}$.

\section{The Rindler horizon\label{sec:rindler}}

In this section we consider the transition rate and the energy density in the closely analogous situation of an inertial observer approaching the Rindler horizon in $(1+1)$-dimensional Minkowski spacetime, 
coupled to a massless scalar field in its Rindler state. 
We also contrast this situation with known results in $(3+1)$-dimensional Minkowski spacetime.

\subsection{$1+1$ Rindler}

We consider $(1+1)$-dimensional Minkowski spacetime with the metric 
\begin{align}
ds^2 = - dt^2 + dx^2 , 
\end{align}
and therein the right-hand-side Rindler wedge, $|t| < x$, 
and therein the geodesic 
\begin{align}
(t,x) = (\tau, \tau_h) , 
\label{eq:Rindler-geodesic}
\end{align}
parametrised by the proper time~$\tau$, 
where $\tau_h$ is a positive constant, 
and the range of $\tau$ within the Rindler wedge is $-\tau_h < \tau < \tau_h$. 
As $\tau\to\tau_h$, the trajectory approaches the future Rindler horizon. 

We take the scalar field to be in the Rindler state, 
for which the positive 
frequencies are defined with respect to the boost Killing vector $x\partial_t + t \partial_x$. 
The Wightman function reads \cite{Birrell:1982ix}
\begin{align}
\mathcal{W}_\text{R}(\mathsf{x},\mathsf{x'}) 
= -\frac{1}{4 \pi} \ln \! \left[(\epsilon+i \Delta u) (\epsilon + i \Delta v) \right] , 
\label{eq:Rindler-wightman}
\end{align}
where the coordinates $(u,v)$ are defined by 
$u = - \ln\bigl(a(x-t)\bigr)$ and $v = \ln\bigl(a(x+t)\bigr)$, 
$\Delta u = u-u'$ and $\Delta v = v-v'$, 
the $\epsilon$-notation specifies the branches of the logarithm as in Section~\ref{seq:quantum-scalar}, 
and $a$ is a positive constant of dimension inverse length that we have included for dimensional consistency. 
In the coordinates $(u,v)$, the geodesic \eqref{eq:Rindler-geodesic} reads 
\begin{subequations}
\begin{align}
u(\tau) &= - \ln\bigl(a(\tau_h-\tau)\bigr) , 
\label{eq:rindler-u-traj}
\\
v(\tau) &= \ln\bigl(a(\tau_h+\tau)\bigr) . 
\end{align}
\end{subequations}

Following the notation of Section~\ref{sec:detector-near}, 
we denote the detector's switch-on moment by $\tau_0$ and switch-off moment by $\tau$, 
where $-\tau_h < \tau_0 < \tau < \tau_h$, and we consider the asymptotic behaviour 
of the transition rate 
$\dot{\mathcal{F}}(\omega,\tau, \tau_0)$ \eqref{Fdot} as $\tau\to\tau_h$, 
with $\tau_0$ fixed. 

Proceeding as in Section~\ref{sec:detector-near}, 
we find that the $\Delta v$-dependent part of 
$\mathcal{W}_\text{R}$ \eqref{eq:Rindler-wightman} remains bounded as 
$\tau\to\tau_h$, whereas, 
by comparison of \eqref{eq:rindler-u-traj} and~\eqref{eq:dot-vtilde-as}, 
the contributions from the $\Delta u$-dependent part obey the same estimates that were found in 
Appendix \ref{app:HRminus-asympt} 
for approaching $H_-^R$ in the Unruh state. We find 
\begin{align}
\dot{\mathcal{F}}_R(\omega,\tau, \tau_0) 
= - \frac{1}{4\pi(\tau_h-\tau)}  
+ 
\frac{1 + o(1)}{2\pi (\tau_h-\tau)\bigl(- \ln(\tau_h-\tau) \bigr)} , 
\label{eq:Fdot-Rindler-final}
\end{align}
where the subscript $R$ refers to the Rindler state. 

The renormalised stress-energy tensor $T_{ab}^{R}$ in the Rindler state can be evaluated by the 
conformal scaling technique as in 
\eqref{Tren-HHI} and~\eqref{Tren-Unruh}, and is well known\cite{Takagi:1986kn}. 
The energy density seen by an observer on the trajectory \eqref{eq:Rindler-geodesic} 
evaluates to 
\begin{align}
\mathcal{E}_{\text{R}}
&= \dot{x}^a \dot{x}^b
T_{ab}^{R}
\notag\\
&= 
- \frac{1}{48\pi} \! 
\left( \frac{1}{{(\tau_h-\tau)}^2} + \frac{1}{{(\tau_h+\tau)}^2} \right) . 
\label{eq:rindler-energydensity}
\end{align}

We see that the divergences in the transition rate \eqref{eq:Fdot-Rindler-final}
and the energy density \eqref{eq:rindler-energydensity} 
are similar to those on approaching the $H_-^R$ branch 
of the Cauchy horizon in the black hole spacetime when the field is in the Unruh state, 
found in Sections \ref{sec:detector-near} and~\ref{sec:stress-energy}, 
including the sign of the divergence and the power law of the divergence. 

\subsection{$3+1$ Rindler}

In $3+1$ spacetime dimensions, an inertial detector approaching the Rindler horizon was analysed in \cite{Louko:2007mu}, taking the field to be in the Rindler state and assuming that the detector's coupling to the field does not include a derivative. In our notation of~\eqref{eq:Rindler-geodesic}, 
the result for the transition rate reads 
\begin{align}
\dot{\mathcal{F}}_{R,3+1}(\omega, \tau, \tau_0) 
& = \frac{1}{8 \pi^2 \tau_h } \Biggl\{\ln\!\left(1 - \frac{\tau}{\tau_h} \right) 
\notag\\
& \hspace{3ex}
+ 2 \ln \!\left[-\ln\!\left(1-\frac{\tau}{\tau_h}\right) +  \mathcal{O}(1)\right] \Biggr\}.
\label{eq:3+1:Fdot-Rindler}
\end{align}
The energy density seen by this inertial observer can be found using 
the Rindler state stress-energy tensor given in~\cite{Takagi:1986kn}, 
with the result 
\begin{align}
\mathcal{E}_{\text{R},3+1}
&= \dot{x}^a \dot{x}^b
T_{ab}^{R,3+1}
\notag\\
&= 
- 
\frac{3\tau_h^2 + \tau^2}{1440\pi^2 {(\tau_h^2 - \tau^2)}^3} . 
\label{eq:rindler-3+1-energydensity}
\end{align}

\subsection{Comparison of $1+1$ and $3+1$}

One might have expected an agreement in the leading divergences of the transition rate $\dot{\mathcal{F}}_R$ \eqref{eq:Fdot-Rindler-final} and $\dot{\mathcal{F}}_{R,3+1}$ \eqref{eq:3+1:Fdot-Rindler}, given the differing short-separation divergences of the Wightman function in $1+1$ and $3+1$ dimensions\cite{Decanini:2005eg}, and the inclusion of the time derivative 
in the coupling in $1+1$ dimensions but not in $3+1$ dimensions. 
Yet the divergences not agree: instead, the leading divergences of $\dot{\mathcal{F}}_R$ agree with the divergences of the $\tau$-derivative of $\dot{\mathcal{F}}_{R,3+1}$. 

These properties of the Rindler state suggest the conjecture that in $3+1$ spacetime dimensions, 
a detector coupled linearly to the value (as opposed to the derivative) of the scalar field, and  approaching a Cauchy horizon, may have a transition rate that diverges only logarithmically in proper time. We leave the investigation of this conjecture to future work.

\section{Conclusions\label{sec:conclusions}}

We have investigated the internal transitions in a space-and-time-localised quantum system on geodesics that approach the Cauchy horizon in a $(1+1)$-dimensional eternal black hole spacetime whose global structure mimics that of the non-extremal Reissner-Nordstr\"om solution. The quantum system was a spatially pointlike Unruh-DeWitt detector, coupled linearly to the proper time derivative of a massless scalar field, which was prepared initially in the HHI state or the Unruh state. Working in first-order perturbation theory, we found that the detector's transition rate generically diverges on approaching the Cauchy horizon, proportionally to the inverse proper time to the horizon. The exception was when the surface gravities of the two horizons are equal: in this case the transition rate remains bounded on all parts of the Cauchy horizon in the HHI state, and on one branch of the Cauchy horizon in the Unruh state. We also saw that these properties of the detector's transition rate have a close qualitative and quantitative similarity with the energy density seen by an observer falling towards the Cauchy horizon. Finally, by comparison with results for the Rindler state and Rindler horizon, we conjectured that a similar but weaker divergence may be present in $3+1$ spacetime dimensions in the transition rate of a detector coupled linearly to the value (rather than to the proper time derivative) of the quantum field. 

That horizons with equal surface gravities emerge as the exceptionally 
regular special case may not be surprising: 
that this case is special was already known from consideration 
of the stress-energy tensor \cite{Hollands:2019whz,Hollands:2020qpe,Klein:2021les}, and similar observations arise with `lukewarm' black holes,  
in which two Killing horizons bound a spacetime region 
in which the Killing vector in question is timelike \cite{Romans:1991nq,Winstanley:2007tf}. 
It is however notable that in the HHI state, 
the overall sign of the leading divergence is determined by which of the two surface gravities is greater, 
in precisely the same way for both the detector's transition rate and for the energy density seen by an observer. 
In comparison, for a detector falling to the singularity of the $(1+1)$-dimensional Schwarzschild spacetime, 
the leading divergence in the detector's transition rate has always a positive sign, 
both in the HHI state and in the Unruh state\cite{Juarez-Aubry:2014jba}. 
This highlights the differences between a Cauchy horizon and a Schwarzschild-type singularity. 

We emphasise that a negative transition rate is not as such physically pathological, given the operational definition of the transition rate in terms of ensembles of ensembles of detectors, switched off at different times\cite{Louko:2007mu,Langlois:2005if}. 
That the integral of the transition rate can diverge to negative infinity may be more disconcerting, 
but this is just an artefact of our passing to the sharp switching limit, 
and dropping the concomitant infinite additive constant from the transition probability. 
If the sharp switching is replaced by a smooth switching, all probabilities are by construction non-negative, 
but disentangling the switching effects from the spacetime effects becomes less transparent, 
as exemplified by a smoothly switched detector that falls in the 
$(3+1)$-dimensional Schwarzschild black hole\cite{Ng:2021enc}. 

Our techniques can be extended to more general spacetimes with horizons, 
such as degenerate horizons\cite{Conroy:2021aow},
or to spacetimes in which spacetime singularities have been resolved by 
nonlinear effects\cite{Ayon-Beato:1999kuh} or by quantum gravity effects\cite{Gambini:2014qta}. 
We leave such extensions subject to future work.

\begin{acknowledgments}
We thank Bernard Kay, Adrian Ottewill and Silke Wein\-furt\-ner for helpful discussions and comments. 
BAJ-A is supported by a CONACYT Postdoctoral Research Fellowship, 
and acknowledges in addition the support of 
CONACYT project 140630 and UNAM-DGAPA-PAPIIT grant IG100120.
JL was supported in part by 
United Kingdom Research and Innovation (UKRI) 
Science and Technology Facilities Council (STFC) grant ST/S002227/1 
``Quantum Sensors for Fundamental Physics'' 
and Theory Consolidated Grant ST/P000703/1. 
\end{acknowledgments}

\section*{Data availability}

Data sharing is not applicable to this article as no new data were created or analyzed in this study. 

\section*{Conflict of interest}

The authors have no conflicts to disclose.

\appendix

\section{$\dot{\mathcal{F}}$ near $H_-^L$\label{app:HLminus-asympt}}

In this appendix we perform the estimates that lead from \eqref{eq:genstatelim}
to the results stated in Section \ref{sec:detector-near} 
for the asymptotics of $\dot{\mathcal{F}}$ near~$H_-^L$.

\subsection{Preliminaries} 

Recall that the geodesic is by assumption in Region~II, 
where $(U,V) \in \BbbR_+ \times \BbbR_+$ and $(\tilde{u},\tilde{v}) \in \BbbR \times \BbbR$. 
The geodesic is an integral curve of the system~\eqref{eqmotion-interior}, 
or equivalently~\eqref{eq:tildedots}. 
Differentiating \eqref{eq:tildedots} gives 
\begin{subequations}
\label{eq:tildeddots}
\begin{align}
\ddot {\tilde u} &  = - \tfrac12 {\dot{\tilde u}}^2 F'(r) , 
\\
\ddot {\tilde v} &  = - \tfrac12 {\dot{\tilde v}}^2 F'(r) , 
\end{align}
\end{subequations}
which will be useful below. 

From 
\eqref{eq:HHI-wightman}
and 
\eqref{eq:U-wightman}
we have, for $\tau'<\tau$, 
\begin{subequations}
\begin{align}
\mathcal{W}_\text{H}(\tau,\tau') 
& = - \frac{1}{4\pi}
\ln[U(\tau) - U(\tau')] 
\notag\\
& \hspace{3ex} 
- \frac{1}{4\pi}
\ln[V(\tau) - V(\tau')] 
- \frac{i}{4} , 
\\
\mathcal{W}_\text{U}(\tau,\tau') 
& = - \frac{1}{4\pi}
\ln[U(\tau) - U(\tau')] 
\notag\\
& \hspace{3ex} 
- \frac{1}{4\pi}
\ln[v(\tau) - v(\tau')] 
- \frac{i}{4} , 
\end{align}
\end{subequations}
and hence 
\begin{subequations}
\label{eq:wboth-firstder}
\begin{align}
\partial_\tau \mathcal{W}_\text{H}(\tau,\tau') 
& = - \frac{\kappa_+}{4\pi}
\times 
\frac{\dot{\tilde u}(\tau)}{1 - U(\tau')/U(\tau)} 
\notag\\
& \hspace{3ex} 
- \frac{\kappa_+}{4\pi}
\times 
\frac{\dot{\tilde v}(\tau)}{1 - V(\tau')/V(\tau)}  , 
\\
\partial_\tau \mathcal{W}_\text{U}(\tau,\tau') 
& = - \frac{\kappa_+}{4\pi}
\times 
\frac{\dot{\tilde u}(\tau)}{1 - U(\tau')/U(\tau)} 
\notag\\
& \hspace{3ex} 
- \frac{1}{4\pi}
\times 
\frac{\dot{\tilde v}(\tau)}{{\tilde v}(\tau) - {\tilde v}(\tau')} . 
\label{eq:wU-firstder}
\end{align}
\end{subequations}
Expanding \eqref{eq:wboth-firstder} in $\tau-\tau'$, 
using \eqref{eq:tildedots} and~\eqref{eq:tildeddots}, we find 
\begin{subequations}
\label{eq:wboth-firstder-limit}
\begin{align}
& \lim_{\tau'\to\tau} 
\left( \partial_\tau \mathcal{W}_\text{H}(\tau,\tau') + \frac{1}{2\pi(\tau-\tau')} \right) 
\notag\\
& \hspace{3ex} = \frac{\left[ \dot{\tilde u}(\tau) + \dot{\tilde v}(\tau) \right] 
\! \left[ F'\bigl(r(\tau)\bigr) - 2 \kappa_+ \right]}{16\pi} , 
\\
& \lim_{\tau'\to\tau} 
\left( \partial_\tau \mathcal{W}_\text{U}(\tau,\tau') + \frac{1}{2\pi(\tau-\tau')} \right) 
\notag\\
& \hspace{3ex} = \frac{\dot{\tilde u}(\tau) \! \left[ F'\bigl(r(\tau)\bigr) - 2 \kappa_+ \right] 
+ \dot{\tilde v}(\tau) F'\bigl(r(\tau)\bigr)}{16\pi} . 
\label{eq:wU-firstder-limit}
\end{align}
\end{subequations}

\subsection{Geodesics approaching $H^L_-$} 

We now specialise to the geodesics approaching~$H^L_-$, 
which are those with $E>0$. 
It follows from \eqref{eq:rdot-interior} that $r(\tau) \to r_-$ 
smoothly and with a nonvanishing derivative as $\tau\to\tau_h$. 
As $\tau\to\tau_h$, ${\tilde v}$ increases smoothly to a finite value and 
$V$ increases smoothly to a finite positive value, 
whereas 
${\tilde u}\to\infty$, logarithmically in $\tau_h-\tau$, 
and $U\to\infty$, as an inverse power-law in $\tau_h-\tau$. 

Given the above observations, and setting $\tau' = \tau_0$ in~\eqref{eq:wboth-firstder}, 
we see that the only contributions to 
$\dot{\mathcal{F}}(\omega, \tau, \tau_0)$ in \eqref{eq:genstatelim} that may be potentially unbounded 
as $\tau\to\tau_h$ come from those terms in 
\eqref{eq:wboth-firstder}
and 
\eqref{eq:wboth-firstder-limit}
that involve~$\dot{\tilde u}$, and these terms are the same for the HHI and Unruh states. 
We may hence drop the reference to the state.

\subsection{First boundary term in \eqref{eq:genstatelim}\label{app:sub1}}

Let $B_1(\tau,\tau_0,\omega)$ denote the first term on the right-hand side of~\eqref{eq:genstatelim}, 
\begin{align}
B_1(\tau,\tau_0,\omega) \doteq -2 \cos(\omega \Delta \tau) \partial_\tau \mathcal{W}(\tau, \tau_0) . 
\label{eq:B1-def}
\end{align}

For $\mathcal{W}(\tau,\tau_0)$, 
\eqref{eq:wboth-firstder} with $\tau'=\tau_0$ gives 
\begin{align}
\partial_\tau \mathcal{W}(\tau,\tau_0) 
 = 
-\frac{\kappa_+\dot{\tilde{u}}(\tau)}{4 \pi}  \left[ 1 + \mathcal{O} \! \left( \frac{U(\tau_0)}{U(\tau)} \right) 
\right] 
+ \mathcal{O}(1).
\label{app:Est1}
\end{align}
To estimate~$\dot{\tilde{u}}$, we write 
\begin{align}
F(r) = - 2 \kappa_- (r-r_-) + \mathcal{O}\bigl((r-r_-)^2 \bigr) , 
\label{eq:F-rminus-expansion}
\end{align}
by which \eqref{eq:dot-utilde} gives 
\begin{align}
\dot{\tilde{u}} = \frac{E}{\kappa_- (r-r_-)} + \mathcal{O}(1), 
\end{align}
and \eqref{eq:rdot-interior} gives 
\begin{align}
r - r_-  = E (\tau_h-\tau) + \mathcal{O}\bigl((\tau_h-\tau)^2 \bigr) , 
\label{eq:r-rminus-expansion}
\end{align}
whence 
\begin{align}
\dot{\tilde{u}}(\tau) = \frac{1}{ \kappa_-(\tau_h-\tau)} + \mathcal{O}(1).
\label{eq:dot-utilde-as}
\end{align}
For $U(\tau_0)/U(\tau)$, we have 
\begin{align}
\frac{U(\tau_0)}{U(\tau)} & = \exp\bigl[\kappa_+\bigl({\tilde u}(\tau_0) - {\tilde u}(\tau)\bigr)\bigr]
\notag\\
& = \exp \left[- \kappa_+\int_{\tau_0}^\tau \! d\tau' \, \dot{\tilde{u}}(\tau') \right] 
\notag\\
& = \exp \left[- \frac{\kappa_+}{\kappa_-} \int_{\tau_0}^\tau \! d\tau' \left(\frac{1}{\tau_h-\tau'} + \mathcal{O}(1) \right) \right] \nonumber \\
& = \exp \left[\frac{\kappa_+}{\kappa_-} \ln \! \left(\frac{\tau_h-\tau}{\tau_h-\tau_0} \right) + \mathcal{O}(1) \right] 
\notag\\
&= \left(\frac{\tau_h - \tau}{\tau_h-\tau_0} \right)^{\kappa_+/\kappa_-} \times e^{\mathcal{O}(1)}
\notag\\
&= \mathcal{O}\bigl((\tau_h-\tau)^{\kappa_+/\kappa_-} \bigr) . 
\label{app:Uestimate}
\end{align}

Collecting, we have 
\begin{align}
&B_1(\tau,\tau_0,\omega)
= \frac{\cos (\omega \Delta \tau) }{2\pi (\tau_h-\tau)}  
\notag\\
& \hspace{3ex}
\times \! \left( \frac{\kappa_+}{\kappa_-} + \mathcal{O}(\tau_h - \tau) + \mathcal{O}\bigl((\tau_h-\tau)^{\kappa_+/\kappa_-}\bigr) \right) . 
\label{app:boundaryresult1}
\end{align}

\subsection{Second boundary term in \eqref{eq:genstatelim}\label{app:sub2}}

Let $B_2(\tau,\tau_0,\omega)$ denote the second term on the right-hand side of~\eqref{eq:genstatelim}. 
From \eqref{eq:wboth-firstder-limit} we have 
\begin{align}
B_2(\tau,\tau_0,\omega) =  \frac{\dot{\tilde u}(\tau) \! \left[ F'\bigl(r(\tau)\bigr) - 2 \kappa_+ \right]}{8\pi} 
+ \mathcal{O}(1) . 
\end{align}
Using \eqref{eq:F-rminus-expansion}, \eqref{eq:r-rminus-expansion} and \eqref{eq:dot-utilde-as}
we obtain 
\begin{align}
B_2(\tau,\tau_0,\omega) =  - \frac{1}{4\pi(\tau_h-\tau)} \! \left( \frac{\kappa_+}{\kappa_-} + 1 \right) 
+ \mathcal{O}(1) . 
\label{app:boundaryresult2}
\end{align}

\subsection{Integral term in \eqref{eq:genstatelim}\label{app:sub3}}

Let $J(\tau,\tau_0,\omega)$ denote the integral term on the right-hand side of~\eqref{eq:genstatelim}. 
As the contribution from the term 
${(\tau-\tau')}^{-1}\sin\bigl(\omega(\tau-\tau')\bigr)$ in the integrand is~$\mathcal{O}(1)$, 
using \eqref{eq:wboth-firstder} gives 
\begin{equation}
J(\omega,\tau,\tau_0) 
= 
\frac{\kappa_+ \omega}{2\pi} \, \dot{\tilde u}(\tau) I(\omega,\tau,\tau_0)  + \mathcal{O}(1) , 
\label{eq:J-est1}
\end{equation}
where 
\begin{equation}
I(\omega,\tau,\tau_0) \doteq \int_{\tau_0}^{\tau} \! d \tau' \, \frac{ \sin \bigl(\omega(\tau-\tau')\bigr)}{1-U(\tau')/U(\tau)} . 
\label{app:I}
\end{equation}
We shall show that the leading contribution in $I(\omega,\tau,\tau_0)$ 
is $\mathcal{O}(1)$ and evaluate this contribution. 

Changing variables in \eqref{app:I} by $\tau' = \tau-s$ gives 
\begin{subequations}
\label{app:split}
\begin{align}
I(\omega, \tau, \tau_0) & = 
I_1(\omega, \tau, \tau_0)
+ 
I_2(\omega, \tau, \tau_0) ,
\\
I_1(\omega, \tau, \tau_0) & = 
\int_{0}^{\tau_h-\tau_0} \! d s \, \frac{ \sin(\omega s)}{1-U(\tau-s)/U(\tau)} , 
\label{app:split1}
\\
I_2(\omega, \tau, \tau_0) & = 
- \int^{\tau_h-\tau_0}_{\tau-\tau_0} \! d s \, 
\frac{ \sin(\omega s)}{1-U(\tau-s)/U(\tau)} , 
\label{app:split2}
\end{align}
\end{subequations}
where in $I_1$ \eqref{app:split1} the upper limit of integration has been extended 
from $\tau-\tau_0$ to $\tau_h-\tau_0$, 
and $I_2$ \eqref{app:split2} has been introduced to compensate for this. 
The extension is well defined provided 
$\tau_h-\tau$ is so small that the detector's 
trajectory may be extended from proper time $\tau_0$ 
backwards to proper time $\tau_0 - (\tau_h-\tau)$, 
still in Region~II, 
which we may assume without loss of generality. 

For $I_2$~\eqref{app:split2}, we have 
\begin{align}
|I_2| & \le 
\int^{\tau_h-\tau_0}_{\tau-\tau_0} \frac{ds}{1-U(\tau-s)/U(\tau)} 
\notag\\
& \le 
\int^{\tau_h-\tau_0}_{\tau-\tau_0} \frac{ds}{1-U(\tau_0)/U(\tau)} 
\notag\\
& = \frac{\tau_h-\tau}{1-U(\tau_0)/U(\tau)} = \mathcal{O}(\tau_h-\tau) , 
\label{eq:I2-est}
\end{align}
using that $U$ is a positive and increasing function of its argument. 

For $I_1$~\eqref{app:split1}, 
we shall show below in Section \ref{app:lemma1} that 
\begin{align}
I_1(\omega, \tau, \tau_0) 
& = 
\frac{1 - \cos\bigl(\omega (\tau_h-\tau_0)\bigr)}{\omega}
\ \ + o(1) 
\notag\\
& = 
\frac{1 - \cos(\omega \Delta\tau)}{\omega}
\ \ + o(1) , 
\label{app:split1-limit}
\end{align}
where in the second equality $\tau_h$ has been replaced by $\tau$ 
at the expense of an $\mathcal{O}(\tau_h-\tau)$ error, 
covered by the~$o(1)$ term. 

Collecting, and using~\eqref{eq:dot-utilde-as}, we have 
\begin{equation}
J(\omega,\tau,\tau_0) 
= 
\frac{(\kappa_+/\kappa_-)}{2\pi (\tau_h - \tau)}  \bigl[1 - \cos(\omega\Delta\tau) + o(1) \bigr] , 
\label{eq:J-result}
\end{equation}

\subsection{Combining}

Adding \eqref{app:boundaryresult1}, \eqref{app:boundaryresult2} and \eqref{eq:J-result} gives 
\begin{align}
\frac{1}{4\pi (\tau_h - \tau)} \left( \frac{\kappa_+}{\kappa_-}  - 1 + o(1) \right) , 
\label{eq:app-evaluation-final}
\end{align}
which is the result shown in \eqref{eq:fdot-final-hhi-LR} and~\eqref{eq:fdot-final-u-left}.

\subsection{Interlude: estimate for $I_1$~\eqref{app:split1}\label{app:lemma1}}

We now establish the estimate \eqref{app:split1-limit} for 
$I_1$~\eqref{app:split1}. 

Recall that $U$ is a positive and strictly increasing function of the proper time along the geodesic, 
with the asymptotics \eqref{app:Uestimate} near the Cauchy horizon. 
It follows that we can write  
\begin{align}
I_1(\omega, \tau, \tau_0) = 
K(\tau_h-\tau,\tau_h-\tau_0) , 
\label{eq:I1-ito-K}
\end{align}
where  
\begin{align}
K(\epsilon,m,\omega) 
\doteq 
\int_{0}^m \! d s \, 
\frac{ \sin(\omega s)}{1-\cfrac{H(\epsilon+s)}{H(\epsilon)}} , 
\label{eq:K-def}
\end{align}
such that $H(x) = x^{-A}\tilde{H}(x)$, 
$A = \kappa_+/\kappa_-$ is a positive constant, 
$\tilde{H}$ is a smooth positive function on $[0, R]$ for some $R>0$, 
$H'<0$ on $(0, R]$, 
$0<m <R$ and $0<\epsilon<R-m$. We consider $m$ and $\omega$ as parameters and wish to find 
$\lim_{\epsilon\to0} K(\epsilon,m,\omega)$. 

For fixed positive~$s$, the ratio $H(\epsilon+s)/H(\epsilon)$ in \eqref{eq:K-def} tends to zero as $\epsilon\to0$. If the limit $\epsilon\to0$ can be taken under the integral, we hence have 
\begin{align}
\lim_{\epsilon\to0}
K(\epsilon,m,\omega) 
& = 
\int_{0}^{m} \! d s \, 
\sin(\omega s) 
\notag\\
& 
= 
\frac{1 - \cos(\omega m)}{\omega} , 
\label{eq:K-limit}
\end{align}
from which the first equality in \eqref{app:split1-limit} follows. 
We shall show that taking the limit under the integral is justified by dominated convergence. 

From $H(x) = x^{-A}\tilde{H}(x)$ and the properties of $\tilde{H}$ 
it follows that there exists 
$\epsilon_1 \in (0, R/2)$ such that $\partial^2_x \ln\bigl(H(x)\bigr) >0$ for 
$0<x < 2\epsilon_1$. This implies that 
we have
\begin{align}
\frac{\partial}{\partial\epsilon} \! \left[ 1-\frac{H(\epsilon+s)}{H(\epsilon)} \right] 
= 
\frac{H(\epsilon+s)}{H(\epsilon)}
\! \left[ \frac{H'(\epsilon)}{H(\epsilon)} - \frac{H'(\epsilon+s)}{H(\epsilon+s)}\right] 
< 0
\label{eq:K-est-aux-ineq}
\end{align}
for $s \in (0,\epsilon_1]$ and $\epsilon \in (0,\epsilon_1)$. 

Taking from now on $\epsilon \in (0,\epsilon_1)$, we split \eqref{eq:K-def} as 
\begin{subequations}
\begin{align}
K(\epsilon,m,\omega) 
& = 
K_<(\epsilon,m,\omega) 
+ 
K_>(\epsilon,m,\omega) , 
\\
K_<(\epsilon,m,\omega) 
& = 
\int_{0}^{\epsilon_1} \! d s \, 
\frac{ \sin(\omega s)}{1-\cfrac{H(\epsilon+s)}{H(\epsilon)}} , 
\label{eq:Ksmall-def}
\\
K_>(\epsilon,m,\omega) 
& = 
\int_{\epsilon_1}^m \! d s \, 
\frac{ \sin(\omega s)}{1-\cfrac{H(\epsilon+s)}{H(\epsilon)}} . 
\label{eq:Klarge-def}
\end{align}
\end{subequations}
By~\eqref{eq:K-est-aux-ineq}, 
the integrand in \eqref{eq:Ksmall-def} is bounded in absolute value by 
\begin{align}
\left|
\frac{ \sin(\omega s)}{1-\cfrac{H(\epsilon+s)}{H(\epsilon)}}
\right| 
\le 
\frac{|\sin(\omega s)|}{1-\cfrac{H(\epsilon_1+s)}{H(\epsilon_1)}} , 
\end{align}
which is independent of $\epsilon$ and integrable over $s \in (0,\epsilon_1)$. 
The integrand in 
\eqref{eq:Klarge-def} is bounded in absolute value by 
\begin{align}
\left|
\frac{ \sin(\omega s)}{1-\cfrac{H(\epsilon+s)}{H(\epsilon)}}
\right| 
\le 
\frac{1}{1-\cfrac{H(\epsilon+\epsilon_1)}{H(\epsilon)}} 
\le 
\frac{1}{1-\cfrac{H(2\epsilon_1)}{H(\epsilon_1)}} , 
\label{eq:Klarge-dominant}
\end{align}
where the first equality comes because $H$ is decreasing and the 
last inequality comes by~\eqref{eq:K-est-aux-ineq}. 
The last expression in \eqref{eq:Klarge-dominant} is independent of 
$\epsilon$ and integrable over $s \in (\epsilon_1, m)$. 
This completes the dominated convergence argument.

\subsection{Special case $\kappa_+ = \kappa_-$}
\label{app:sub-equalkappas}

We now consider the special case $\kappa_+ = \kappa_-$, in which the leading term in 
\eqref{eq:app-evaluation-final} vanishes. 
We show that the transition rate remains bounded as $\tau\to\tau_h$. 

The contributions from the parts involving $\dot{\tilde v}$ in \eqref{eq:wboth-firstder}
and \eqref{eq:wboth-firstder-limit} remain bounded as $\tau\to\tau_h$. 
The $\mathcal{O}$-terms in $B_1(\tau,\tau_0,\omega)$ \eqref{app:boundaryresult1} 
now combine to $\mathcal{O}(\tau_h-\tau)$. 
What needs a better estimate is $I_1(\omega,\tau,\tau_0)$, given by 
\eqref{eq:I1-ito-K} and~\eqref{eq:K-def}, now with $A=1$. 

In \eqref{eq:K-def}, setting $A=1$ and isolating the leading behaviour gives 
\begin{subequations}
\begin{align}
K(\epsilon,m,\omega) 
& = 
K_0(m,\omega) + \epsilon K_1(\epsilon,m,\omega) , 
\\
K_0(m,\omega) 
& = 
\int_{0}^m \! d s \, 
\sin(\omega s) 
= 
\frac{1 - \cos(\omega m)}{\omega} , 
\\
K_1(\epsilon,m,\omega) 
& = \frac{1}{\tilde{H}(\epsilon)}
\int_{0}^m \! d s \, 
\frac{\sin(\omega s)}{s} 
\times 
\frac{s}{g(\epsilon+s) - g(\epsilon)} , 
\end{align}
\end{subequations}
where $g(x) \doteq x/\tilde{H}(x)$ for $x>0$ and $g(0) \doteq 0$. 
As $g$ is smooth and satisfies $g'(x)>0$ for $x \in [0,R]$, there exists a positive constant~$k_1$, 
independent of~$\epsilon$, such that $k_1 s \le g(\epsilon+s) - g(\epsilon)$ for $s\in[0,m]$ and sufficiently small~$\epsilon$. Dominated convergence hence implies
\begin{align}
\lim_{\epsilon\to0}
K_1(\epsilon,m,\omega) 
= \frac{1}{{\tilde H}(0)}
\int_{0}^m \! d s \, 
\frac{\sin(\omega s)}{s} {\tilde H}(s) , 
\end{align}
which is finite. 

Combining, it follows that $\dot{\mathcal{F}}(\omega,\tau,\tau_0)$ remains bounded as $\tau\to\tau_0$.

\section{$\dot{\mathcal{F}}$ near $H_-^R$\label{app:HRminus-asympt}}

In this appendix we perform the estimates that lead from \eqref{eq:genstatelim}
to the results stated in Section \ref{sec:detector-near} 
for the asymptotics of $\dot{\mathcal{F}}$ near~$H_-^R$. 

The geodesics in question are those with $E<0$ in \eqref{eqmotion-interior} and~\eqref{eq:tildedots}. 

As the HHI state is invariant under the right-left reflection, $(U,V) \mapsto (V,U)$, 
the result for the HHI state is the same whether the Cauchy horizon branch is $H_-^P$ or~$H_-^F$. 
This gives the result shown in~\eqref{eq:fdot-final-hhi-LR}. 

For the Unruh state, the contributions from the parts in 
\eqref{eq:wU-firstder}
and 
\eqref{eq:wU-firstder-limit}
that involve $\dot{\tilde{u}}$ are smooth as $\tau\to\tau_0$. 
We need to consider the contributions from the parts that involve~$\dot{\tilde{v}}$.

\subsection{First boundary term in \eqref{eq:genstatelim}\label{app:sub1r}}

Consider the first term on the right-hand side of~\eqref{eq:genstatelim}, 
given by $B_1(\tau,\tau_0,\omega)$~\eqref{eq:B1-def}. 
For $\mathcal{W}(\tau,\tau_0)$, 
\eqref{eq:wU-firstder} with $\tau'=\tau_0$ gives 
\begin{align}
\partial_\tau \mathcal{W}_\text{U}(\tau,\tau_0) 
= 
- \frac{1}{4\pi}
\times 
\frac{\dot{\tilde v}(\tau)}{{\tilde v}(\tau) - {\tilde v}(\tau_0)} 
+\mathcal{O}(1) . 
\label{eq:rest1}
\end{align}
Proceeding as with \eqref{eq:dot-utilde-as} gives 
\begin{align}
\dot{\tilde{v}}(\tau) = \frac{1}{ \kappa_-(\tau_h-\tau)} + \mathcal{O}(1).
\label{eq:dot-vtilde-as}
\end{align}
Hence \begin{align}
B_1(\tau,\tau_0,\omega)
= \frac{\cos (\omega \Delta \tau) }{2\pi (\tau_h-\tau)\bigl(- \ln(\tau_h-\tau) + \mathcal{O}(1) \bigr)} . 
\label{app:right-boundaryresult1}
\end{align}

\subsection{Second boundary term in \eqref{eq:genstatelim}\label{app:sub2r}}

For the second term on the right-hand side of~\eqref{eq:genstatelim}, 
$B_2(\tau,\tau_0,\omega)$, \eqref{eq:wU-firstder-limit} gives 
\begin{align}
B_2(\tau,\tau_0,\omega) = 
\frac{\dot{\tilde v}(\tau) F'\bigl(r(\tau)\bigr)}{16\pi}
+ \mathcal{O}(1) . 
\end{align}
Proceeding as with \eqref{app:boundaryresult2} gives 
\begin{align}
B_2(\tau,\tau_0,\omega) =  - \frac{1}{4\pi(\tau_h-\tau)}  
+ \mathcal{O}(1) . 
\label{app:right-boundaryresult2}
\end{align}

\subsection{Integral term in \eqref{eq:genstatelim}\label{app:sub3r}}

Let $J(\tau,\tau_0,\omega)$ again denote the integral term on the right-hand side of~\eqref{eq:genstatelim}. 
Proceeding as with \eqref{eq:J-est1}, using \eqref{eq:wU-firstder} gives 
\begin{equation}
J(\omega,\tau,\tau_0) 
= 
\frac{\omega \dot{\tilde v}(\tau)}{2\pi {\tilde v}(\tau)} \tilde{I}(\omega,\tau,\tau_0)  + \mathcal{O}(1) , 
\end{equation}
where 
\begin{equation}
\tilde{I}(\omega,\tau,\tau_0) \doteq \int_{\tau_0}^{\tau} \! d \tau' \, \frac{ \sin \bigl(\omega(\tau-\tau')\bigr)}{1-{\tilde v}(\tau')/{\tilde v}(\tau)} , 
\end{equation}
and we are assuming $\tau$ to be so close to $\tau_h$ that ${\tilde v}(\tau)$ is positive. 
Proceeding as in~\eqref{app:split}, we have 
\begin{subequations}
\label{app:split-tilde}
\begin{align}
\tilde{I}(\omega, \tau, \tau_0) & = 
\tilde{I}_1(\omega, \tau, \tau_0)
+ 
\tilde{I}_2(\omega, \tau, \tau_0) ,
\\
\tilde{I}_1(\omega, \tau, \tau_0) & = 
\int_{0}^{\tau_h-\tau_0} \! d s \, \frac{ \sin(\omega s)}{1-{\tilde v}(\tau-s)/{\tilde v}(\tau)} , 
\label{app:split1-tilde}
\\
\tilde{I}_2(\omega, \tau, \tau_0) & = 
- \int^{\tau_h-\tau_0}_{\tau-\tau_0} \! d s \, 
\frac{ \sin(\omega s)}{1-{\tilde v}(\tau-s)/{\tilde v}(\tau)} . 
\label{app:split2-tilde}
\end{align}
\end{subequations}

For $\tilde{I}_2$~\eqref{app:split2-tilde}, 
proceeding as in \eqref{eq:I2-est} shows that $\tilde{I}_2 = \mathcal{O}(\tau_h-\tau)$.  

For $\tilde{I}_1$ \eqref{app:split1-tilde}, we may proceed as in Section~\ref{app:lemma1}, 
using now the asymptotic behaviour of ${\tilde v}$ obtained from \eqref{eq:dot-vtilde-as} 
to show that the limit $\tau\to\tau_h$ limit can be taken under the integral, 
with the result 
\begin{align}
\tilde{I}_1(\omega, \tau, \tau_0) 
& = 
\frac{1 - \cos\bigl(\omega (\tau_h-\tau_0)\bigr)}{\omega}
\ \ + o(1) 
\notag\\
& = 
\frac{1 - \cos(\omega \Delta\tau)}{\omega}
\ \ + o(1) . 
\label{app:split1-tilde-limit}
\end{align}
Hence 
\begin{align}
J(\omega,\tau,\tau_0) 
=
\frac{1 - \cos (\omega \Delta \tau) + o(1)}{2\pi (\tau_h-\tau)\bigl(- \ln(\tau_h-\tau) + \mathcal{O}(1) \bigr)} . 
\label{eq:app-J-righthor}
\end{align}

\subsection{Combining}

Adding \eqref{app:right-boundaryresult1}, \eqref{app:right-boundaryresult2}
and 
\eqref{eq:app-J-righthor} gives 
\begin{align}
- \frac{1}{4\pi(\tau_h-\tau)}  
+ 
\frac{1 + o(1)}{2\pi (\tau_h-\tau)\bigl(- \ln(\tau_h-\tau) \bigr)} , 
\label{eq:J-unruh-righthor}
\end{align}
which gives the result shown in~\eqref{eq:fdot-final-u-right}.

\section{$\dot{\mathcal{F}}$ near the Cauchy horizon bifurcation point\label{app:Hbif}}

In this appendix we perform the estimates that lead from \eqref{eq:genstatelim}
to the results stated in Section \ref{sec:detector-near} 
for the asymptotics of $\dot{\mathcal{F}}$ near the Cauchy horizon bifurcation point. 

The geodesics are those with $E=0$ in \eqref{eqmotion-interior} and~\eqref{eq:tildedots}. 
Proceeding as in Appendix~\ref{app:HLminus-asympt}, 
we find that \eqref{eq:dot-utilde-as} holds but the error term can be improved to 
\begin{align}
\dot{\tilde{u}}(\tau) = \frac{\bigl[ 1 + p\bigl((\tau_h - \tau)^2\bigr) \bigr]}{ \kappa_-(\tau_h-\tau)} , 
\label{eq:bif-dot-utilde-as}
\end{align}
where $p$ is a smooth function of a nonnegative argument such that $p(x) = \mathcal{O}(x)$, and similarly for $\dot{\tilde{v}}$. It follows that the estimate \eqref{app:Uestimate} for $U(\tau_0)/U(\tau)$ improves to 
\begin{align}
\frac{U(\tau_0)}{U(\tau)} 
&= \left(\frac{\tau_h - \tau}{\tau_h-\tau_0} \right)^{\kappa_+/\kappa_-} \times 
q\bigl((\tau_h - \tau)^2\bigr) , 
\label{app:bif-Uestimate}
\end{align}
where $q$ is a smooth positive function of a non-negative argument. 

We now need to consider in 
\eqref{eq:wboth-firstder}
and 
\eqref{eq:wboth-firstder-limit}
both the terms that involve $\dot{\tilde u}$ and the terms that involve~$\dot{\tilde v}$. 
For the terms that involve~$\dot{\tilde u}$, all the estimates given in 
Appendix \ref{app:HLminus-asympt} still hold. 
For the terms that involve~$\dot{\tilde v}$, all the estimates given in 
Appendix \ref{app:HRminus-asympt} still hold for the Unruh state, while for the HHI state the outcome is the same as with the terms involving~$\dot{\tilde u}$, 
by the left-right symmetry of the state. Combining these observations leads to 
\eqref{eq:fdot-final-hhi-0}~and~\eqref{eq:fdot-final-u-bif}. 

In the Unruh state, the case $\kappa_+/\kappa_- = 2$ is exceptional because the leading term in \eqref{eq:fdot-final-u-bif} 
vanishes due to cancellations. To find the leading term in this case, we need a better estimate for 
$K(\epsilon,m,\omega)$ \eqref{eq:K-def} with $A=2$. 
It is here that we need the improved estimate~\eqref{app:bif-Uestimate}. 

In \eqref{eq:K-def}, setting $A=2$ and isolating the leading behaviour gives now 
\begin{subequations}
\begin{align}
K(\epsilon,m,\omega) 
& = 
K_0(m,\omega) + \epsilon K_1(\epsilon,m,\omega) , 
\\
K_0(m,\omega) 
& = 
\int_{0}^m \! d s \, 
\sin(\omega s) 
= 
\frac{1 - \cos(\omega m)}{\omega} , 
\\
K_1(\epsilon,m,\omega) 
& = \frac{1}{\tilde{H}(\epsilon)}
\int_{0}^m \! d s \, 
\frac{\sin(\omega s)}{s} 
\times 
\frac{\epsilon s}{\tilde{g}\bigl((\epsilon+s)^2\bigr) - \tilde{g}\bigl(\epsilon^2\bigr)} , 
\label{eq:bif-K1-def}
\end{align}
\end{subequations}
where $\tilde{g}(x) \doteq x/\tilde{H}(\sqrt{x}\,)$ for $x>0$ and $\tilde{g}(0) \doteq 0$. 
By the improved estimate~\eqref{app:bif-Uestimate}, 
$\tilde{g}$ is smooth and satisfies $\tilde{g}'(x)>0$ for $x \in [0,\sqrt{R}]$. 
There thus exists a positive constant~$k_1$, 
independent of~$\epsilon$, such that $2 k_1 \epsilon s \le k_1 \bigl((\epsilon+s)^2 - \epsilon^2\bigr) \le \tilde{g}\bigl((\epsilon+s)^2\bigr) - \tilde{g}\bigl(\epsilon^2\bigr)$ for $s\in[0,m]$ and sufficiently small~$\epsilon$. This provides a dominated convergence bound that justifies taking the $\epsilon\to0$ limit of $K_1(\epsilon,m,\omega)$ \eqref{eq:bif-K1-def} under the integral, and the limit is zero. Hence 
$K(\epsilon,m,\omega) = 
K_0(m,\omega) + \epsilon o(1)$ as $\epsilon\to0$. 
This and \eqref{eq:J-unruh-righthor} give~\eqref{eq:fdot-final-u-bif-special}.

\end{document}